 \documentclass[final,authoryear,1p,11pt]{elsarticle}

\usepackage{graphicx}
\usepackage{epstopdf}
\usepackage{booktabs}
\usepackage{natbib}
\usepackage{fancyhdr}
\usepackage{amsmath}
\usepackage{amsfonts}
\usepackage{multirow}
\usepackage{lscape}
\usepackage{bbm}
\usepackage{color}
\usepackage{hyperref}
\usepackage{color,soul}
\sethlcolor{green}
\usepackage{ifthen}
\usepackage[latin1]{inputenc}
\usepackage{titletoc}
\usepackage{eso-pic}
\definecolor{123}{rgb}{.9,.9,.9}
\hypersetup{
colorlinks=false,
citecolor=blue,
linkbordercolor={1 1 1}, 
citebordercolor={1 1 1},
urlbordercolor={1 1 1}
}

\usepackage{amssymb}
 \usepackage{amsthm}
 \usepackage{xcolor}

\begin{document}

\begin{frontmatter}

\title{Forecasting the term structure of crude oil futures prices with neural networks\tnoteref{label1}}

\author[ies,utia]{Jozef Barun\'{i}k\corref{cor2}} \ead{barunik@utia.cas.cz}
\author[ies]{Barbora Malinsk\'{a}}
\address[ies]{Institute of Economic Studies, Charles University, Opletalova 21, 110 00, Prague,  CR}
\address[utia]{Institute of Information Theory and Automation, The Czech Academy of Sciences, Pod Vodarenskou Vezi 4, 182 00, Prague, Czech Republic}

\tnotetext[label1]{Support from the Czech Science Foundation under the P402/12/G097 DYME -- ``Dynamic Models in Economics" project is gratefully acknowledged. The research leading to these results has received funding from the European Union's Seventh Framework Programme (FP7/2007-2013) under grant agreement No. FP7-SSH- 612955 (FinMaP).}

\begin{abstract}
The paper contributes to the rare literature modeling term structure of crude oil markets. We explain term structure of crude oil prices using dynamic Nelson-Siegel model, and propose to forecast them with the generalized regression framework based on neural networks. The newly proposed framework is empirically tested on 24 years of crude oil futures prices covering several important recessions and crisis periods. We find 1-month, 3-month, 6-month and 12-month-ahead forecasts obtained from focused time-delay neural network to be significantly more accurate than forecasts from other benchmark models. The proposed forecasting strategy produces the lowest errors across all times to maturity. 
\end{abstract}
\end{frontmatter}

\section{Introduction}

Modeling and forecasting term structures of commodity markets is attractive from academic perspective and valuable for producers, speculators, and risk managers. Generally, term structure illustrates expectations about future development of the corresponding market. Notwithstanding the high importance, there is almost no relevant literature forecasting commodity term structures. In this paper, we introduce a novel framework for forecasting term structure of crude oil futures prices. We propose to couple dynamic neural networks with Nelson-Siegel model to obtain precise forecasts of the crude oil futures prices. 

Crude oil is essential to world economies from the industrial perspective as it is vital input of production and its price is driven by distinct demand and supply shocks. Shifts in the price of oil are driven to different extents by aggregate or precautionary demand related to market anxieties about the availability of future oil supplies.  As demand of crude oil, which is not dependent as much on price as on income \citep{hamilton2009}, continues to rise and supply is probable to decline (due to nature of crude oil as limited resource), literature agrees about highly volatile and hence uncertain future development of crude oil prices \citep{pan2009}. Main reasons for crude oil market being one of the most volatile in the world are rising demand and supply strongly dependent on behavior of politically and economically unstable countries, crude oil demand and production heavily correlated with occurrence of exogenous events such as military conflicts and natural catastrophes, and presence of speculators \citep{buyuksahin2011}. 

With crude oil futures market being one of the most developed markets according to the trading volumes, understanding the behavior of its term structure becomes even more important. Nevertheless, literature modeling and forecasting term structure of petroleum markets is rather scarce (see \cite{lautier2005} for review). Similarly to the interest rate models, there are two approaches of modeling term structure in petroleum commodities. Spot price being a natural candidate for state variable in one-factor model is modeled as geometric Brownian motion \citep{brennan1985}, or mean-reverting process \citep{schwartz1997}. Later, researchers started to consider convenience yield as a second state variable in a two factor model \citep{schwartz1997}. Alternatively, \cite{gabillon1991} employs long-term price as the second state variable. While both approaches assume constant interest rate, which implies that future spot price and forward prices are the same, \cite{cortazar2003} developed the three factor model.

A relatively fresh new surge of literature explaining the commodity futures prices uses the approach of \cite{diebold2006}, originally introduced to model yield curves. Motivated by similarities of stylized facts between commodity markets and interest rate markets, dynamic Nelson-Siegel model is a natural candidate for this task. Among few, \cite{karstanje2015common} examine the comovement of factors driving commodity futures curves and their shapes by adopting the framework of the dynamic Nelson-Siegel model \citep{diebold2006}. Joint dynamics of factors driving commodity futures curves using multiple-regime framework is further studied by \cite{nomikos2015petroleum}. \cite{almansour2014convenience} model the futures term structure of crude oil and natural gas markets with switching regimes, and \cite{heidorn2015impact} regress futures curve factors extracted from dynamic Nelson-Siegel model on fundamental and financial traders. While dynamic Nelson-Siegel model explains the dynamics of factors underlying term structure of commodity prices, literature is silent about the future predictions with only exception of \cite{gronborg2015nalyzing}. In their original work, \cite{diebold2006} propose to use a simple autoregressive time series models to successfully forecast the dynamics of term structure factors, and hence prices in the interest rates market. We hypothesize, that factors in commodity markets may contain further nonlinear dependencies, which need to be modeled in order to obtain precise forecasts. Therefore, application of more general methods which do not require restrictive assumptions about the underlying structure of factors is appropriate.  

A natural candidate for the forecasting task are neural networks, which can be viewed as a generalized non-linear regression tool. Concisely, neural networks are semi-parametric non-linear models, which are able to approximate any reasonable function \citep{Haykin2007,Hornik1989359}. Whereas the number of models using machine learning is rapidly growing in the academic literature, applications in energy markets are very limited. While several works use neural networks in energy forecasting \citep{fan2008,yu2008,xiong2013,jammazi2012,papadimitriou2014forecasting,barunik2014coupling}, we are the first to employ the approach in forecasting of term structures. 

The contribution of this work is twofold. First, we enhance rare literature studying term structure of commodity prices with new results from the application of dynamic Nelson-Siegel modeling strategy on the crude oil futures markets for long period of 1990 -- 2014. Second, we propose to use time-delay neural network to forecast the term structure factors identified by the dynamic Nelson-Siegel model. Using this framework, we forecast the term structure of crude oil futures prices successfully over the 1-month, 3-month, 6-month and 12-month forecasting horizons.

\section{Data}
\label{sec:data}
\subsection{Raw data}
The data set consists of monthly closing prices of West Texas Intermediate (WTI) futures contracts,\footnote{Available at \url{https://www.quandl.com/c/futures/cme-wti-crude-oil-futures}.} traded on the New York Mercantile Exchange (NYMEX). Each contract expires three trading days prior the 25th calendar day in the month preceding the month of delivery.\footnote{Full specification of WTI futures contracts available on \url{http://www.cmegroup.com/trading/energy/files/en-153_wti_brochure_sr.pdf}} In total, we analyze 396 monthly historical (already delivered) and to-date undelivered contracts -- 12 contracts per each year with delivery months in period starting 1990. Undelivered contracts represented in the dataset are contracts with delivery in November, December 2014 and 24 contracts with delivery in two subsequent years 2015 and 2016. 
 
The main reason for using data starting from 1990 is that the maximum time to maturity for contracts before this date was up to nine months, while later during the period it increased to more than six years. Hence to avoid potentially large risk and inaccuracies stemming from data extrapolation, we consider only data after the year 1990. Choice of the monthly frequency is mainly driven by the fact that contracts with longer time to maturity were traded rather infrequently in the first half of the studied period. In addition, \cite{baumeister2015high} find monthly data to have equal predictive ability to daily data. 

 \begin{table}[t]
 \footnotesize
  \centering
    \begin{tabular}{llccccccc}
    \toprule
    Contract &  & \multicolumn{2}{c}{CLQ2003} & \multicolumn{2}{c}{CLU2003} & \multicolumn{2}{c}{CLV2003} &  \\
    Date &       & Settle & $\tau$  & Settle & $\tau$  & Settle & $\tau$  &  \\
  \midrule    
    28.2.2001 &       & 21,72 & 625   & 21,62 & 646   & \multicolumn{1}{c}{-} & \multicolumn{1}{c}{-} & \multicolumn{1}{c}{} \\
    31.3.2001 &       & 22,76 & 602   & 22,70  & 623   & \multicolumn{1}{c}{-} & \multicolumn{1}{c}{-} & \multicolumn{1}{c}{} \\
    30.4.2001 &       & 23,46 & 582   & 23,35 & 603   & \multicolumn{1}{c}{-} & \multicolumn{1}{c}{-} & \multicolumn{1}{c}{} \\
    31.5.2001 &       & 23,57 & 559   & 23,45 & 580   & 23,33 & 603   &  \\
    \bottomrule
    \end{tabular}
    \caption{Example of future prices and corresponding maturities for contracts traded between 28.2.2001 and 31.5. 2001 for different contracts. CME product code \textit{CL} is used for WTI futures contract, the letters \textit{Q, U}, and \textit{V} denote the delivery in August, September and October.}
  \label{tab:example_monthly}
\end{table}
Table \ref{tab:example_monthly} presents an example of actual data to illustrate the structure and dimension of the dataset. In order to associate each observation of futures price with corresponding time to maturity, it is necessary first to find exact expiry date of each contract. Then, the difference between expiry date and date of observation gives us remaining days to maturity. Table \ref{tab:example_monthly} captures end-of-month futures prices of three different (in this case consecutive) contracts with delivery in August, September and October 2003. For example, at the end of February 2001, CLQ2003 and CLU2003 contracts were traded. On February, 28 2001 it was possible to enter into contract with delivery in August 2003 with futures price USD 21,72 per barrel. Respective time to maturity ($\tau$) was 625 trading days.

\begin{table}[hb]
\footnotesize
  \centering
   \begin{tabular}{lcccccccc}
    \toprule
          & \multicolumn{8}{c}{Days to maturity ($\tau$)} \\
    Date  & 30    & 60    & 90    & 120   & 150   & 180   & 210 & $\ldots$ \\
	\midrule    
    28.2.2001 & 27,48 & 27,36 & 26,99 & 26,60 & 26,21 & 25,84 & 25,48 & $\ldots$ \\
    31.3.2001 & 26,50 & 26,59 & 26,43 & 26,20 & 25,94 & 25,68 & 25,43 & $\ldots$ \\
    30.4.2001 & 28,74 & 28,89 & 28,53 & 28,07 & 27,60 & 27,19 & 26,78 & $\ldots$ \\
    31.5.2001 & 28,49 & 28,42 & 28,14 & 27,78 & 27,38 & 27,00 & 26,59  & $\ldots$ \\
    \bottomrule
    \end{tabular}%
    \caption{Example of reorganized data set to constant time to maturity}
  \label{tab:constmat_monthly}
\end{table}

\subsection{Reorganized data}
After combining the days to maturity with each observed quotation of futures price, the desired form of dataset is a matrix with number of rows equal to number of days included in analysis and number of columns equal to number of analyzed maturities. 

\begin{figure}[ht]
    \centering
    \includegraphics[width=0.8\textwidth]{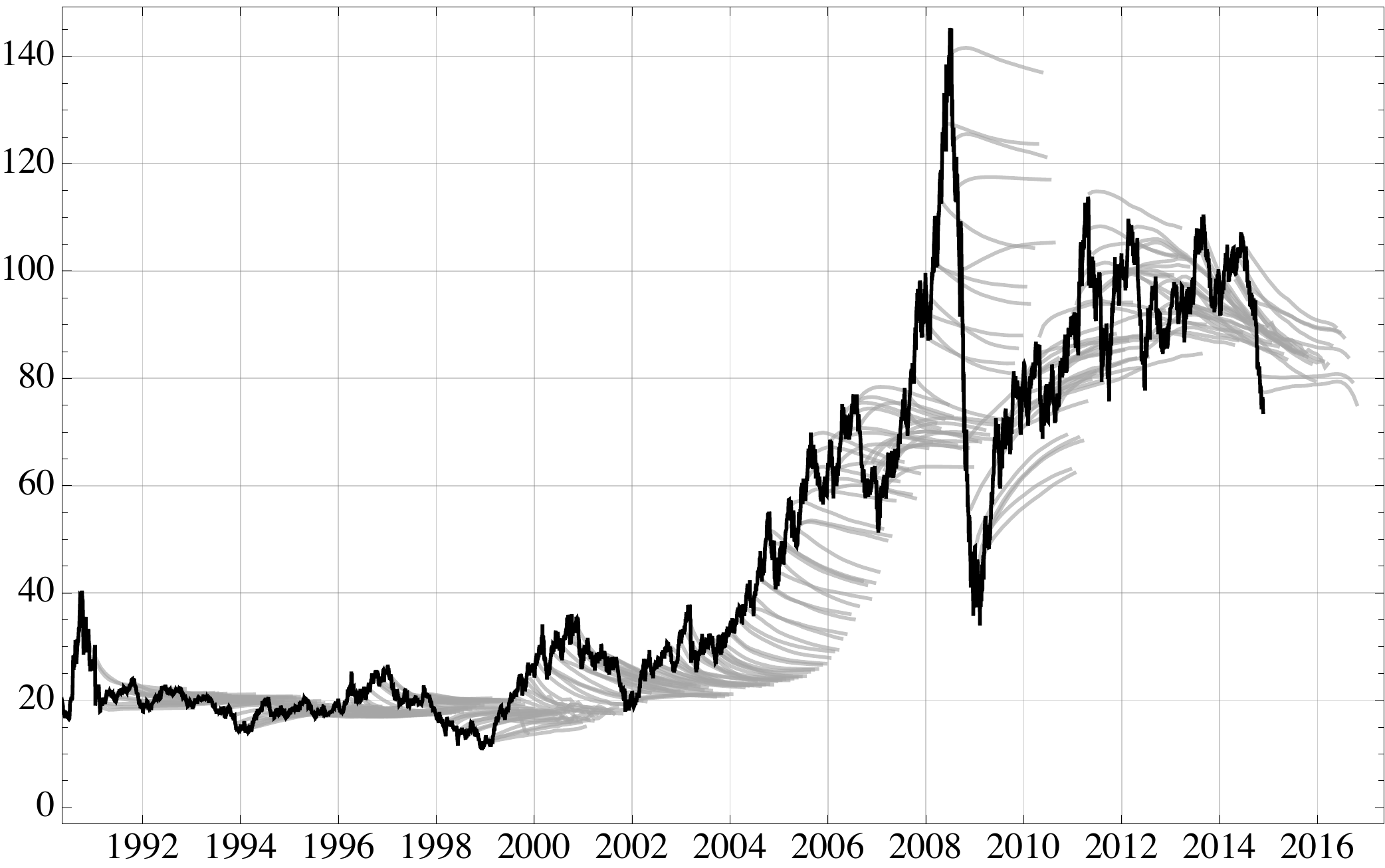}     
    \caption{Reorganized data set: monthly term structures of crude oil futures prices plotted against daily spot prices for the period 1990 -- 2014.}
    \label{fig:term0}
\end{figure}

Time series captured in Table \ref{tab:constmat_monthly} are reorganized constant-maturity futures prices. WTI crude oil futures are delivered and expire with one-month regularity, therefore futures prices with exactly 30, 60, or 90 days to maturity are not traded every day. There are several ways in the literature to interpolate the prices to obtain desired form of the data. \cite{diebold2006} use linear interpolation for constant maturity, while \cite{holton2003} prefers cubic splines interpolation.\footnote{For detailed discussion of interpolation methods for curve construction with applications on yield curve modeling see \cite{hagan2006}.} In our work, we follow the approach of \cite{holton2003}, and use cubic spline interpolation. Figure \ref{fig:term0} illustrates the reorganized constant-maturity futures prices we work with, plot against the daily evolution of the spot price. 

\begin{figure}[ht]
    \centering
    \includegraphics[width=0.45\textwidth]{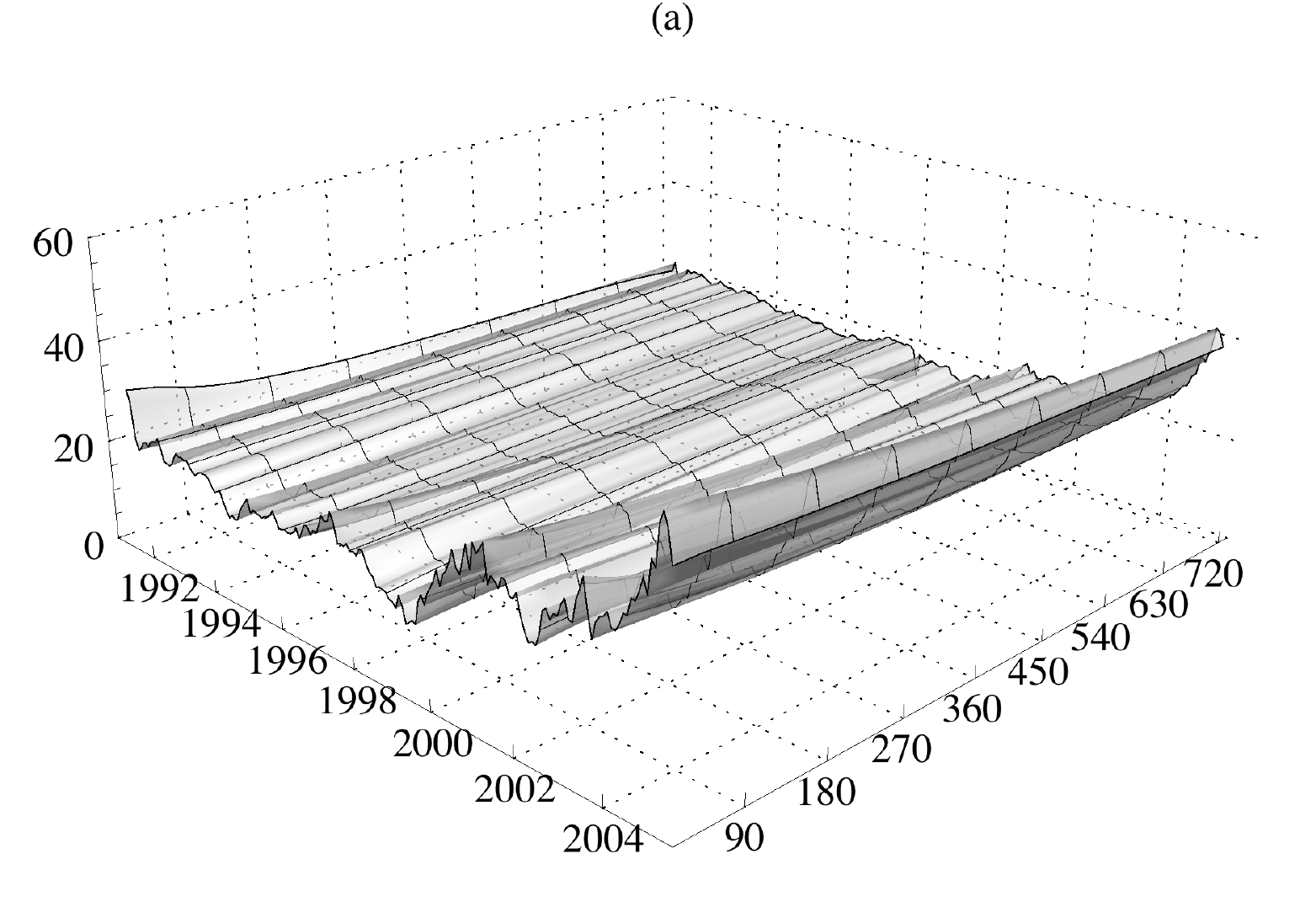}
    \includegraphics[width=0.45\textwidth]{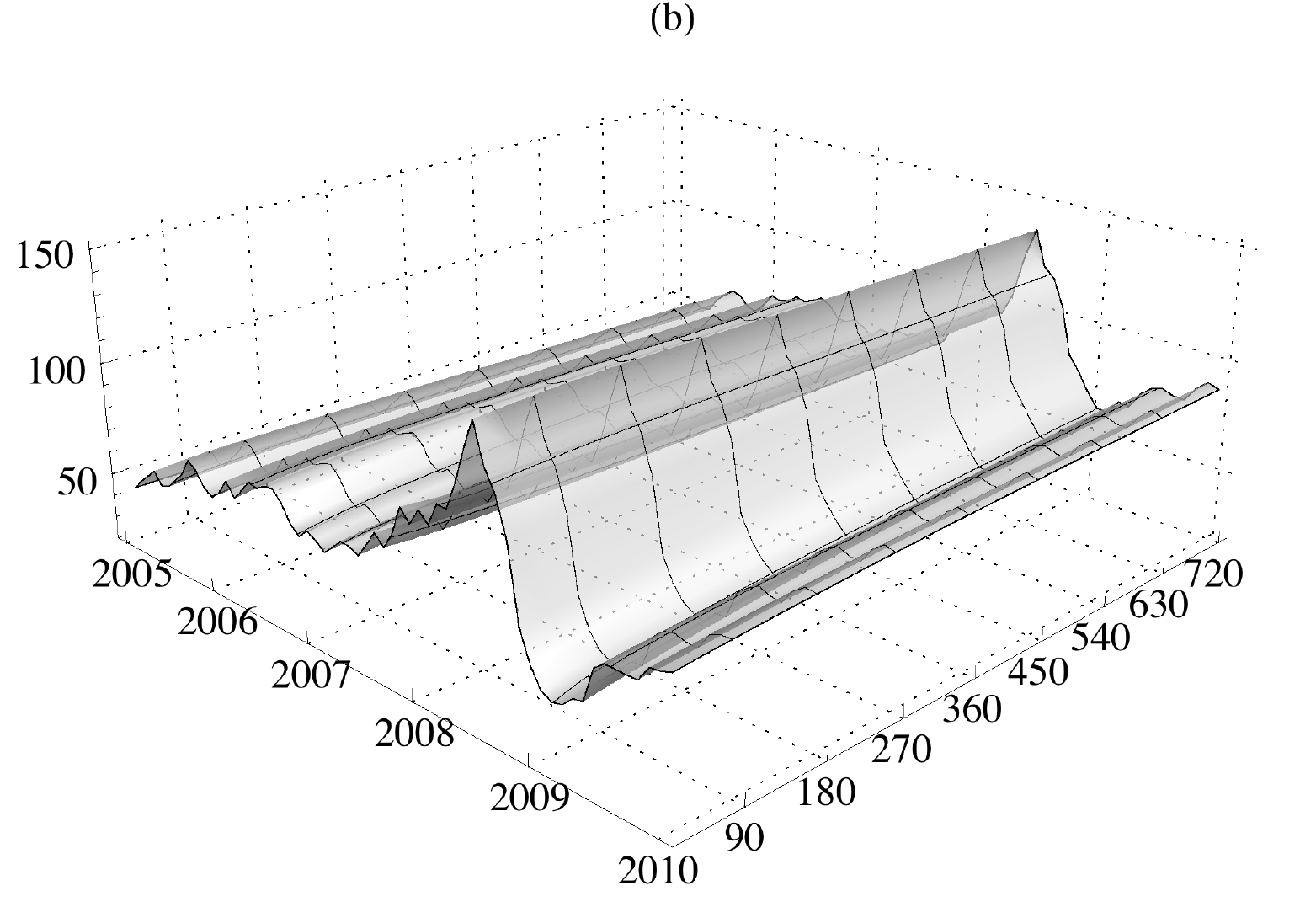}	
    \includegraphics[width=0.45\textwidth]{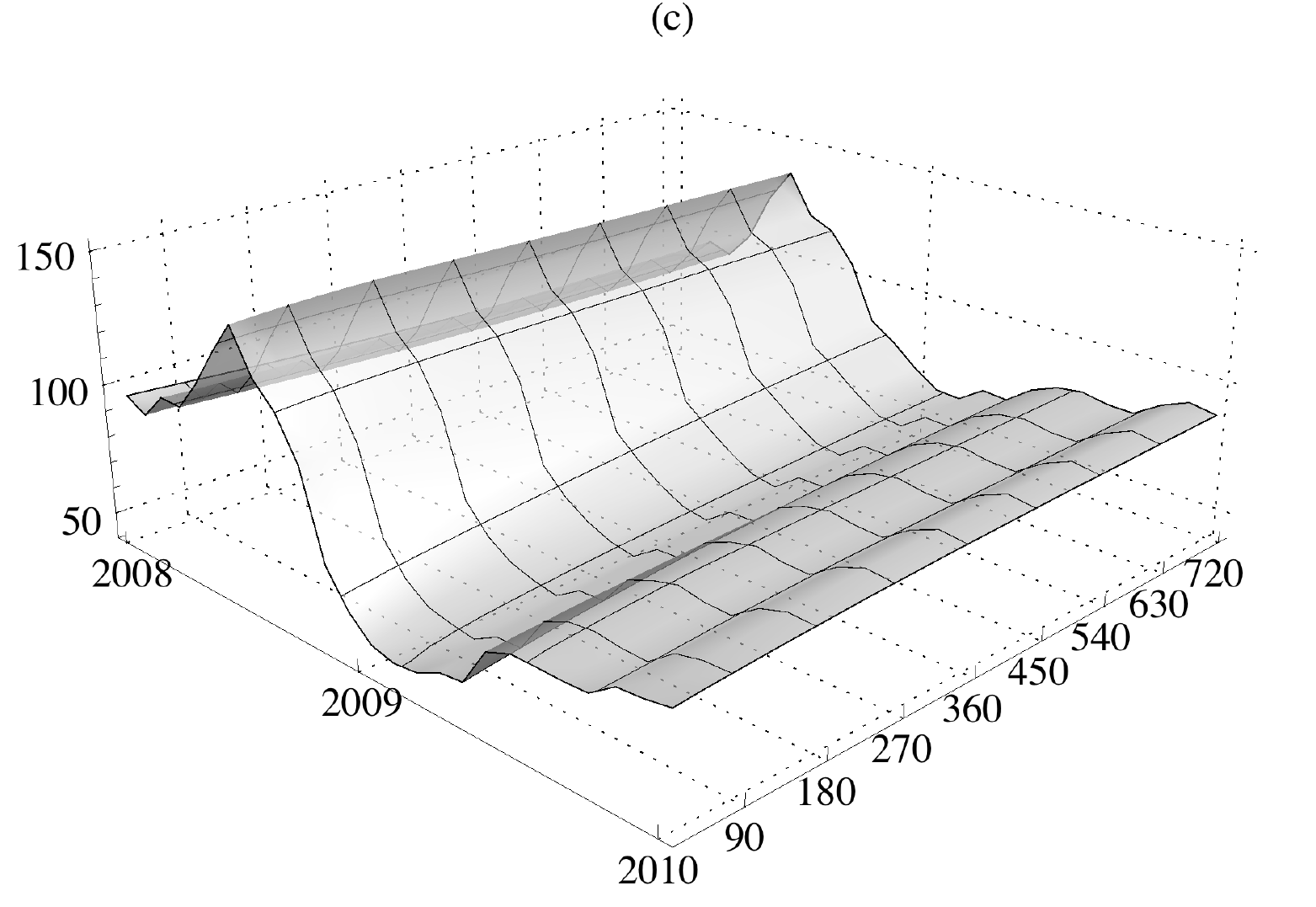}
    \includegraphics[width=0.45\textwidth]{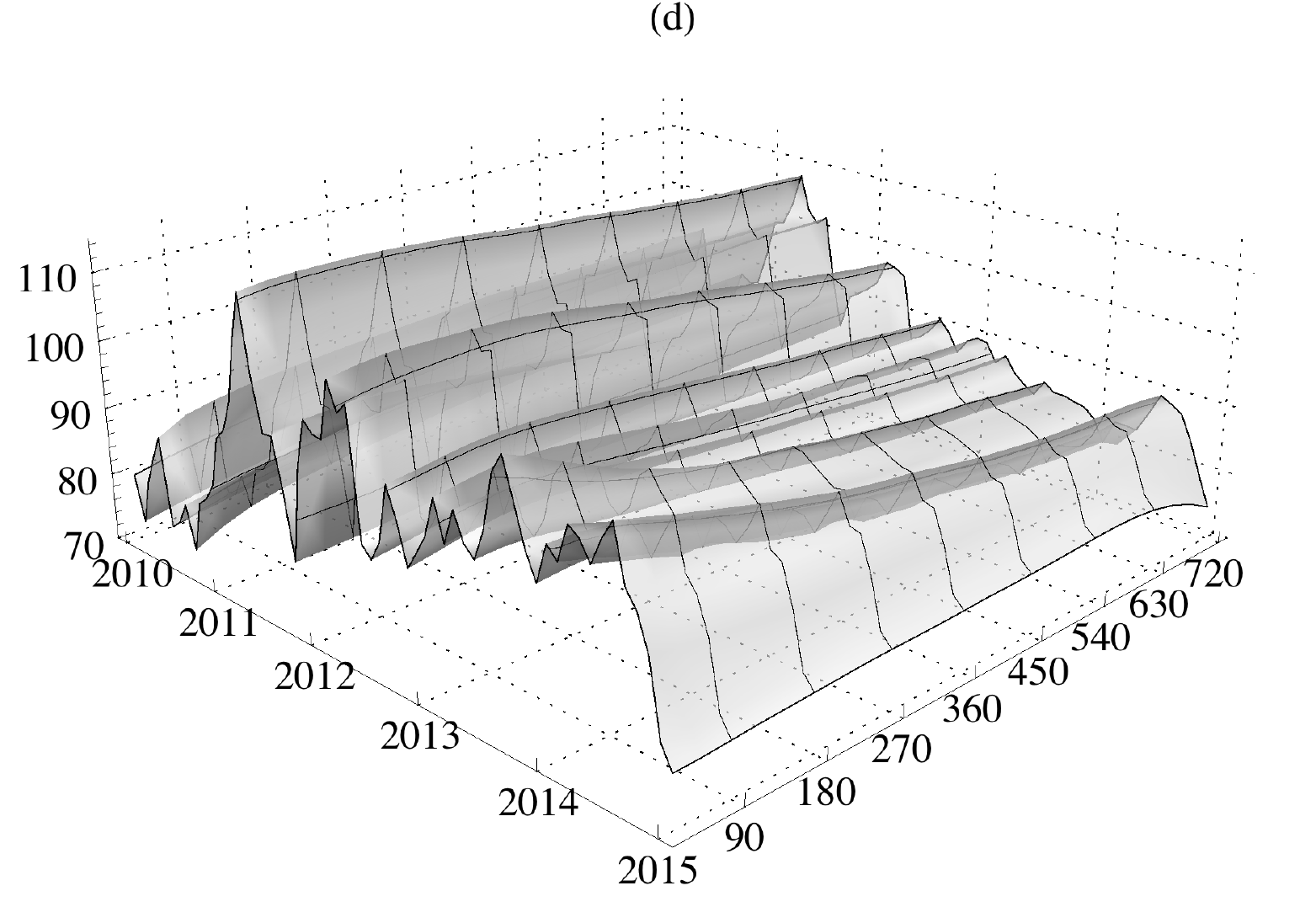}      
    \caption{Term structure of WTI futures prices for the period (a) 1990 -- 2004 , (b) 2005 -- 2009, (c) 2008 -- 2009, and (d) 2010 -- 2014. Dates, days to maturity, and futures prices are on corresponding $\{x,y,z\}$ axes.}
    \label{fig:term}
\end{figure}

Due to a long time span including several turbulent periods, we present term structure in separate periods to better highlight the rich dynamics (Figure \ref{fig:term}).

Figure \ref{fig:term} (a) illustrates the term structure dynamics in period 1990 to 2004. Prices are relatively steady with slight shift downwards during Asian crisis during 1990s with a dramatic change after the year 2000 due to the energy crisis. Some authors attribute increase in futures prices to speculators and sudden shrinkage of oil reserve, while others disprove their arguments \citep{kilian2014,mahadeva2013role}. Upcoming period started with steady dynamics and documented decent increase of futures prices across all maturities form the year 2005 (Figure \ref{fig:term} (b)). The calm period has been interrupted by turbulent one around the year 2008, when crude oil prices exceeded USD 100 per barrel. 

Figure \ref{fig:term} (c) provides more detailed illustration of the rich dynamics during the period. Military conflicts in Nigeria (including oil pipelines attacks), tension between Iran and Israel and consequent fear of oil crisis accelerated rise of oil prices to unprecedented levels. Political unrest in the Middle East together with sharp depreciation of the U.S. Dollar resulted in further frequent and significant horizontal shifts in term structure. Global financial crisis returned WTI term structure back under USD 100 per barrel, and the data exhibit horizontal shift upwards at the end of 2009 driven by complicated political environment in the Middle East -- conflicts in Gaza Strip.
 
Increasing, decreasing or humped shapes of term structure can be observed during the most recent five-year period, as illustrated in Figure \ref{fig:term} (d). WTI term structure experienced strong upward horizontal shift during 2011 caused by political unrest in Egypt or Libya together with the weak U.S. Dollar. Another steep shift upwards in 2012 had also political reason -- danger of closing Strait of Hormuz by Iran as an answer to sanctions against Iran's nuclear programme.\footnote{Approximately 20\% of worldwide traded crude oil passes through the Strait according to the U.S. Energy Information Administration, see \url{http://www.eia.gov/countries/analysisbriefs/World_Oil_Transit_Chokepoints/wotc.pdf}.} Finally, Greek bailout and Chinese economy stimulated by increased money supply contributed to rise of crude oil prices.

\subsection{Stylized facts about term structure}
\label{sec:stylized}

Previous discussion documents many shapes of crude oil futures term structure, which are essentially similar to yield curves of government bonds, although the data are fundamentally different. The similarities has been discussed in detail by \citep{gronborg2015nalyzing}, who compare the five stylized facts about government bonds yield curves \citep{diebold2006} to the stylized facts about crude oil term structures. The discussion is important as we rely on the dynamic Nelson-Siegel approach \citep{diebold2006} for modeling term structure.
 
The main stylized facts about the yield curves are: (1) on average, the yield curve is increasing in time to maturity, and concave, (2) it exhibits various shapes through time -- upward or downward sloping, humped, and inverted humped, (3) the ``near" end of the yield curve is much more volatile than the ``far" end, (4) yield dynamics are persistent, dynamics of spreads are much less persistent, and (5) long rates are more persistent than short rates.

Term structure of crude oil is moreover vulnerable to political decisions and conflicts, hence its shape often changes not only in sense of horizontal shifts, but also in actual shape. To document its ability to exhibit wide variety of shapes we borrow the Figure \ref{fig:shapes} from Section \ref{sec:OLSfit}, documenting four days with different shapes of the analyzed curve as illustrative examples. At the end of November 1990, we can observe smooth decreasing term structure (Figure \ref{fig:shapes} (a)). In May 1999 the curve does not show any smoothness and its behavior is unclear. Figure \ref{fig:shapes} (c) shows nice increasing curve and the most recent example (Figure \ref{fig:shapes} (d)) proves also presence of humped curves in the data.

Probably the most specific feature of crude oil future markets is backwardation.\footnote{Backwardation is a situation when future prices are lower than spot prices.} \cite{hotelling1931} postulates that equilibrium price of non-renewable resources like crude oil, which equals to net marginal revenue, increases  over time at rate of interest. However, key differencing factor between Hotelling's theory and theories of backwardation on crude oil market is uncertainty \citep{litzenberger1995}. As argued by \cite{haubrich2004}, the opposite situation on the market -- contango -- should be present. Futures prices should be above spot prices of crude oil, as opportunity cost equal to interest rate and storage costs make crude oil stocks disadvantageous. Convenience yield justifies occurrence of backwardation on commodity markets. Storing a commodity implies not only costs but also benefits. Convenience yield can be understood as \textit{``... flow of services that accrues to an owner of the physical commodity but not to an owner of a contract for future delivery of the commodity ..."} \citep{brennan1985}. The discounted marginal convenience yields to the the present value then equal backwardation appearing on the market, implying exogenously determined backwardation. One can introduce oil production as a call option to make it endogenous \citep{litzenberger1995}. Alternative explanation was proposed by \cite{lautier2005}, who points out analogy between convenience yield and coupons or dividends linked to bonds and stocks, respectively.

\section{Modeling the term structure}
\label{sec:three}

As motivated by the previous analysis, crude oil term structure is similar to fixed income securities, hence the modeling vehicle can be shared.\footnote{There are simplifying assumptions for the crude oil term structure models -- there are no frictions, taxes, or transaction costs on the market, trading is continuous, lending and borrowing rates are equal, short sale is unconstrained and markets are complete \citep{lautier2005}.} The most successful approach used in the recent literature to model and forecast yield curves has been introduced by \cite{diebold2006}. The model is a dynamic representation of Nelson-Siegel model \citep{nelson1987}, and has been recently used in the crude oil markets successfully by \cite{gronborg2015nalyzing}. Contrary to affine general equilibrium models, which assume concrete functional relationship for yield curve, this class of models does not stem from any theoretical grounds and is based only on parametrization of curve shapes. Generally, models of curve fitting using standard statistical methods perform better in curve fitting and forecasting compared to affine models \citep{steeley2008}.

\subsection{Dynamic Nelson-Siegel model}

For the modeling of term structure of crude oil futures prices, we use the dynamic Nelson-Siegel model \citep{diebold2006}. Choice of this framework is motivated by several aspects. First, other classes of models such as no-arbitrage or affine general equilibrium models fail in forecasting. As \cite{sarker2006} points out, no-arbitrage models focus on cross-section fitting of yield curve at particular point in time, which implies lack of capturing yield curve dynamics by the model. Affine models capture time-series dynamics, but omit proper cross-sectional fit at given time. Second, functional specification of yield curves provided by \cite{nelson1987} is able to model diverse shapes observable on markets. Third, the model provides intuitive parameters, which are straightforward to explain and interpret. Further, \cite{bliss1996} has shown that Nelson-Siegel model outperforms other methods in yield curve estimation, and \cite{diebold2006} show Nelson-Siegel model to be able to replicate stylized facts about yield curves. On the contrary, \cite{duffie1996} concluded that yield curves estimated by affine general equilibrium models, such as Vasicek or CIR, do not conform the behavior observed on markets.  
 
\cite{diebold2006} propose to forecast the yield curve using time series of three yield curve components formulated in Nelson-Siegel model. In this framework, the dynamics of the term structure of crude oil futures prices is described by
\begin{equation}
\label{DNS}
p_t(\tau) = \beta_{0t} + \beta_{1t} \left( \frac{1-e^{-\lambda_t \tau}}{\lambda_t \tau} \right) + \beta_{2t} \left( \frac{1-e^{-\lambda_t \tau}}{\lambda_t \tau} - e^{-\lambda_t \tau} \right)
\end{equation}
where $p_t(\tau)$ is price of crude oil futures at time $t=1,\ldots,T$ with time to maturity $\tau = 30, 60, 90, ..., 720$, and $\beta_{0t}$, $\beta_{1t}$, and $\beta_{2t}$ are interpreted as coefficients on level, slope and curvature factors, respectively. Level factor is long-term component as the values of the factor are constant over whole period and maturities. Slope factor is short-term component, as long as it decays exponentially at rate $\lambda_t$. Finally, curvature factor is referred to as medium-term component, as it increases for medium-term maturities and then decays for the longest maturities.

\begin{figure}[ht]
    \centering
    \includegraphics[width=0.7\textwidth]{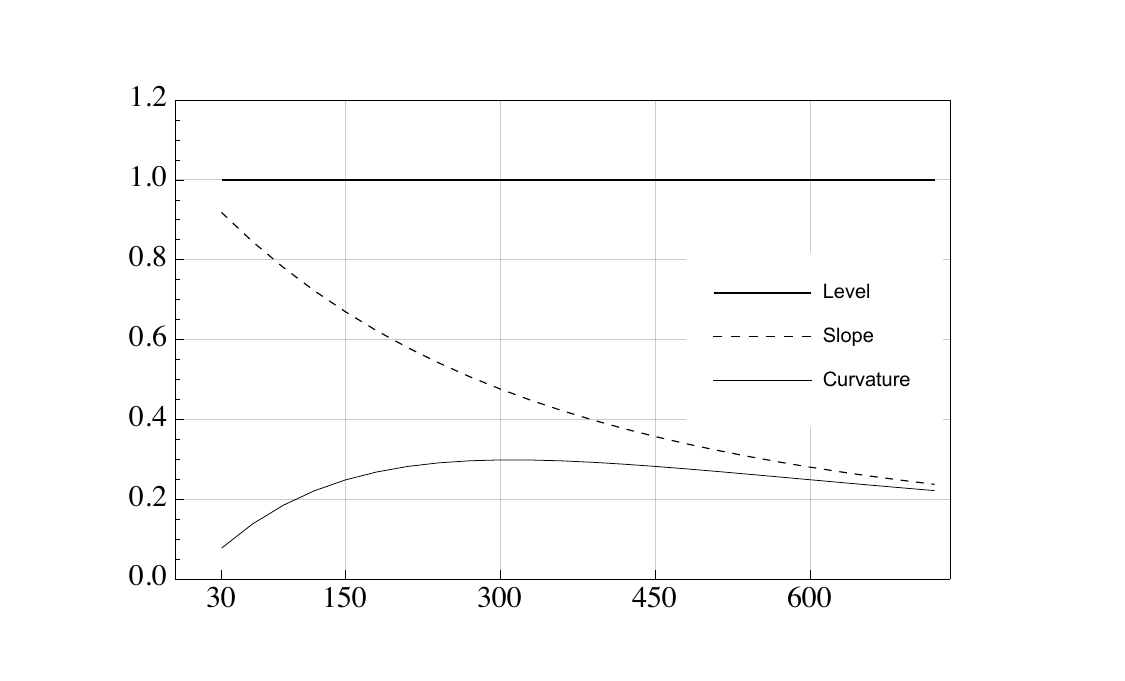}     
    \caption{Loadings of Nelson-Siegel latent factors of term structure}  
    \label{fig:loadings}  
\end{figure}

Figure \ref{fig:loadings} presents estimated loadings of the factors as a function of time to maturity. The plot uses a fixed decay $\lambda_t=\lambda=0.0058$ found empirically in the next section. 

Level factor on the $\beta_{0t}$ is constant for all the maturities, hence impacts futures prices for all maturities evenly. Change to level factor means horizontal shift of term structure, and thus will affect prices at all maturities in the same way. Loading on the slope factor is decreasing from one (zero time to maturity) to zero with maturity going to infinity. Note that Figure \ref{fig:loadings} plots maturities starting from 30 days. Compared to the curvature factor, slope factor is higher for shorter maturities which confirms $\beta_{1t}$ to be rather short-term factor, i.e. affecting prices associated with shorter maturities more. On the contrary, curvature factor converges to zero with time to maturity approaching zero, and infinity, while $\beta_{2t}$ has highest loadings for medium maturities with maximum at time to maturity equal to $1/\lambda$. 

\subsubsection{Decay parameter}

The most important element in Nelson-Siegel class of models is parameter $\lambda_t$ determining exponential decay. Low values of the parameter imply slower decay of the resulting curve and vice versa. Empirically, choice of $\lambda_t$ value represents a trade-off between fitting close and far ends of term structure. Higher values of the parameter result in better fit of the functional form in the case of short maturities. Conversely, lower values improve fit for the longer maturities \citep{diebold2006}. Decay parameter also defines maturity where loading on the medium term curvature factor $\beta_{2t}$ is maximized.  
  \begin{figure}[t]
    \centering
    \includegraphics[width=0.7\textwidth]{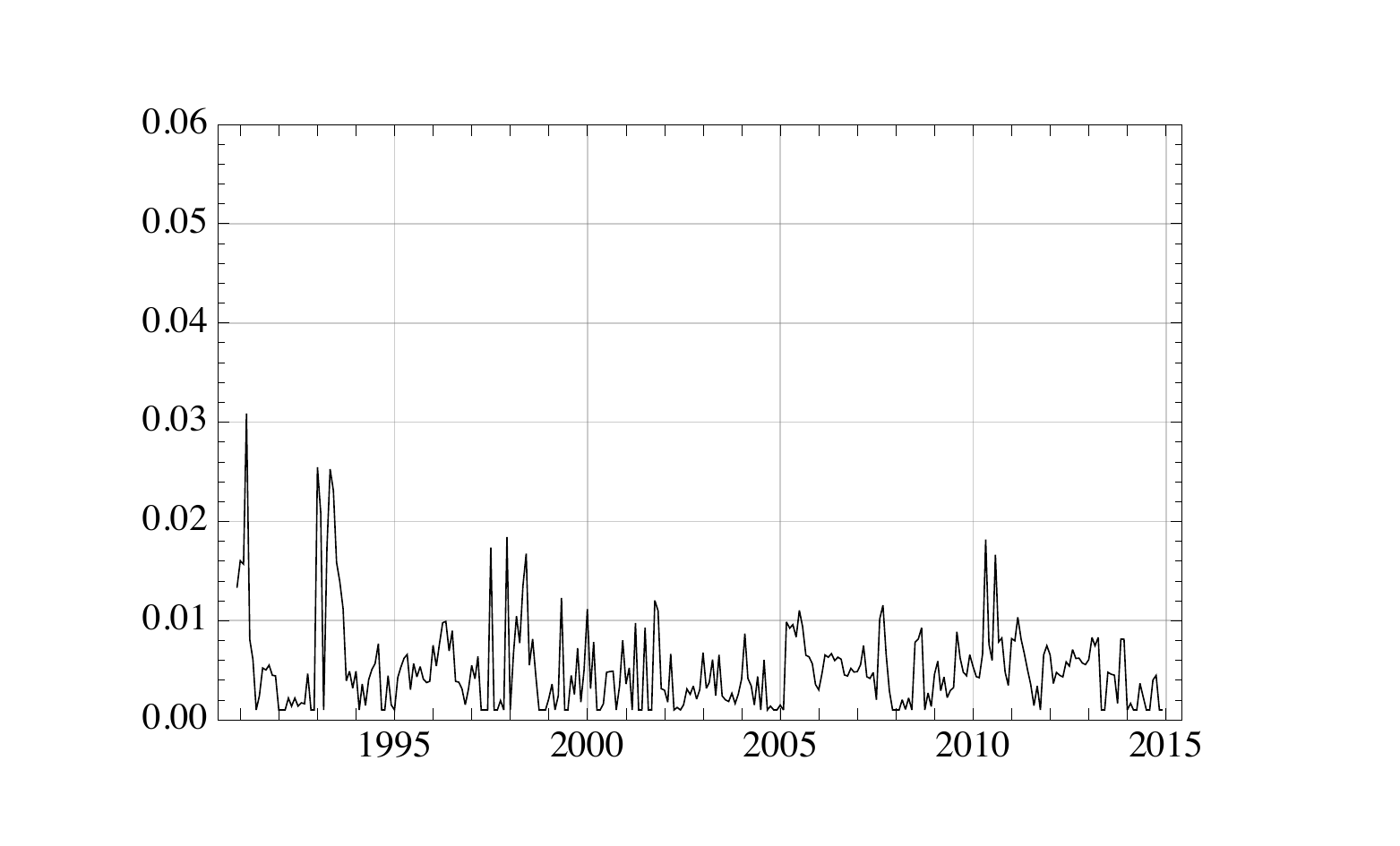}    
    \caption{Time series of $\lambda_t$}
    \label{fig:lambda} 
\end{figure}  

In addition, $\lambda_t$ handling governs actual nature of above defined relationship. If we allow for dynamically evolving $\lambda_t$ over time, we obtain nonlinear problem, which is computationally much more demanding. While authors in the yield curve literature often consider 2-years or 3 -years time to maturity as medium-term maturity, and use this assumption to fix $\lambda_t=\lambda$ for all times $t=1,\ldots,T$, it is infeasible in case of crude oil futures. Literature on crude oil term structure does not provide any well reasoned suggestions about medium-term maturities on oil markets, and there is almost no reference for proper choice of $\lambda$, as modeling term structure of crude oil markets using Nelson-Siegel family models is not fully explored in the literature.
 
A different approach employs nonlinear least squares estimation of all four parameters in Equation \ref{DNS}, i.e. $\beta_{0t}$, $\beta_{1t}$, $\beta_{2t}$ and $\lambda_t$ for all $t$. The main problem of such an approach is, that $\lambda_t$ may be unstable due to unexpected jumps. While the model will fit the data very well, its predictive power deteriorates \citep{vela2013}. 

We find the optimal values of $\lambda_t$ by minimizing sum of squared errors of Nelson-Siegel approximations of WTI futures term structure for each observed point in time. Figure \ref{fig:lambda} illustrates the estimates. To ease the optimization, we restrict the values to correspond maturity between 0 and 1000 days. While $\lambda$ determines reciprocal value to number of days to maturity where medium-term (i.e. curvature) factor is maximized, search for optimal $1/\lambda_t$ outside this interval is superfluous.

We can observe that $\lambda_t$ is unstable for the crude oil futures data showing no clear pattern. Consequently, allowing for dynamic $\lambda_t$ makes successful predictions hardly possible. Therefore, we find single optimal value of $\lambda$ by minimizing sum of squared errors of Nelson-Siegel approximation of WTI term structure over the whole period as
\begin{equation}
\begin{aligned}
& \lambda^* = \underset{\lambda \in \Theta}{\text{argmin}}
& & \sum\limits_{t=1}^{289} \sum\limits_{i=1}^{24} \left(p_t(\tau_i) - \widehat{p}_t(\tau_i;\beta_{0t},\beta_{1t},\beta_{2t},\lambda)\right)^2  \\
\end{aligned}
\end{equation}
where 289 is total number of observed points in time and 24 is number of analyzed constant maturities (from 30 to 720 days). Resulting value of  $\lambda^*=0.0058$, implying reciprocal value of  $1/\lambda^*$ equal to 173.4551 yielding acceptable value of medium term maturity.\footnote{The maximum observed time to maturity in our period reached less than 2000 days = approx. 6 years, which is much less compared to 30 years in case of U.S. yield curve. In such a case authors claim 2 - 3 years to be medium maturity.} Result of the optimization is in line with reviewed literature. \cite{gronborg2015nalyzing}, who analyzed oil futures (although in different period) arrived to $\lambda$ equal to 0.005. 

\subsubsection{Level, slope, and curvature estimates}
\label{sec:OLSfit}
\begin{figure}[t]
    \centering
    \includegraphics[width=0.7\textwidth]{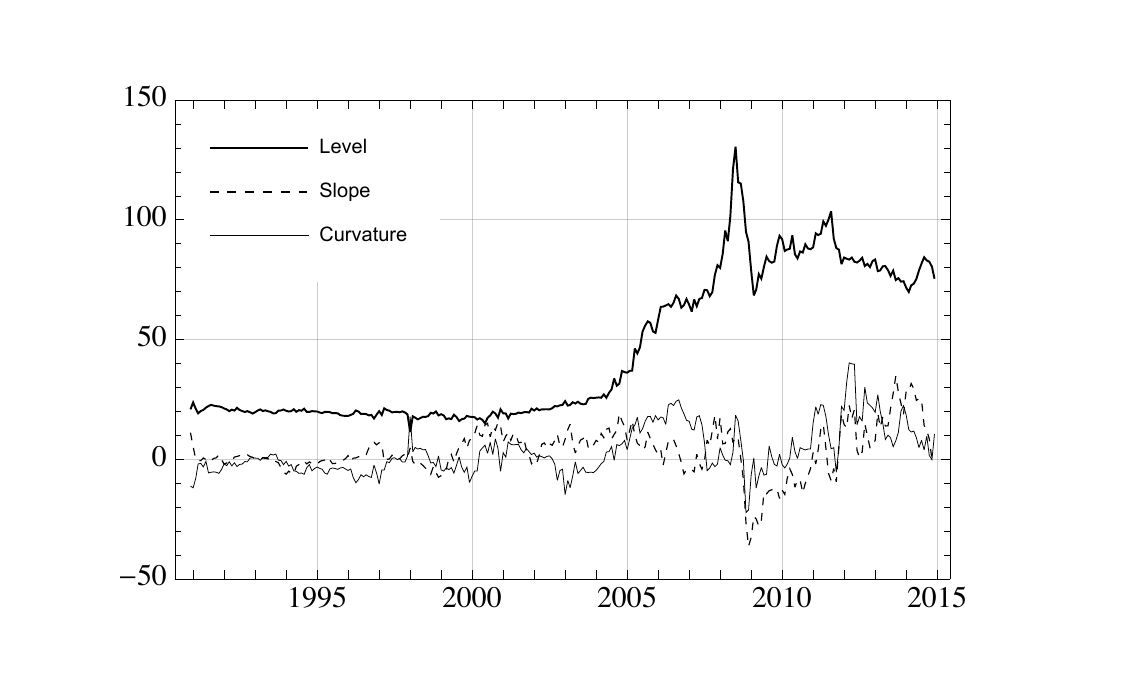}
    \caption{Estimated coefficients from dynamic Nelson-Siegel model of crude oil futures for the period of 1990 -- 2014. Level -- $\beta_{0t}$, slope -- $\beta_{1t}$, and curvature -- $\beta_{2t}$.}
    \label{fig:betas}    
\end{figure} 
Having set the optimal value of $\lambda^*$, we proceed with in-sample estimation of the set of $\beta_t$ coefficients on latent factors. For all times $t$, the parameters are obtained from ordinary least squares (OLS) fit across maturities
\begin{equation}
\label{OLS}
\begin{aligned}
\underset{\beta_{0}, \beta_{1}, \beta_2}{\text{min}}
& & \sum\limits_{i=1}^{24} \left( p_t(\tau_i) - \beta_0 - \beta_1 \left( \frac{1-e^{-\lambda^* \tau_i}}{\lambda^* \tau_i} \right) - \beta_2 \left( \frac{1-e^{-\lambda^* \tau_i}}{\lambda^* \tau_i} - e^{-\lambda^* \tau_i} \right) \right)^2 \\
\end{aligned}
\end{equation}
where  $p_t(\tau_i)$ is WTI futures price at time $t$ with time to maturity $\tau_i$. This procedure results in obtaining time series of three $\beta$-coefficients, with length of 289 values.
 
Estimates of $\beta_t$ coefficients are plotted in Figure \ref{fig:betas}. At first glance, behavior of $\beta_{0t}$ - the level coefficient - attracts attention. Increasing level coefficient over whole observed period corresponds to general increase of crude oil prices. Slope and curvature coefficients seem to be in general more stable. Slope factor fluctuates around zero in the first part of the sample, while it becomes positive until 2008, meaning that resulting term structure is downward sloping. After 2008, slope coefficient jumps to large negative values and remains negative for following two years implying upward sloping term structure. Most recent period from 2011 is characterized by positive values which implies decreasing term structure.
\begin{figure}[t]
    \centering
    \includegraphics[width=1.0\textwidth]{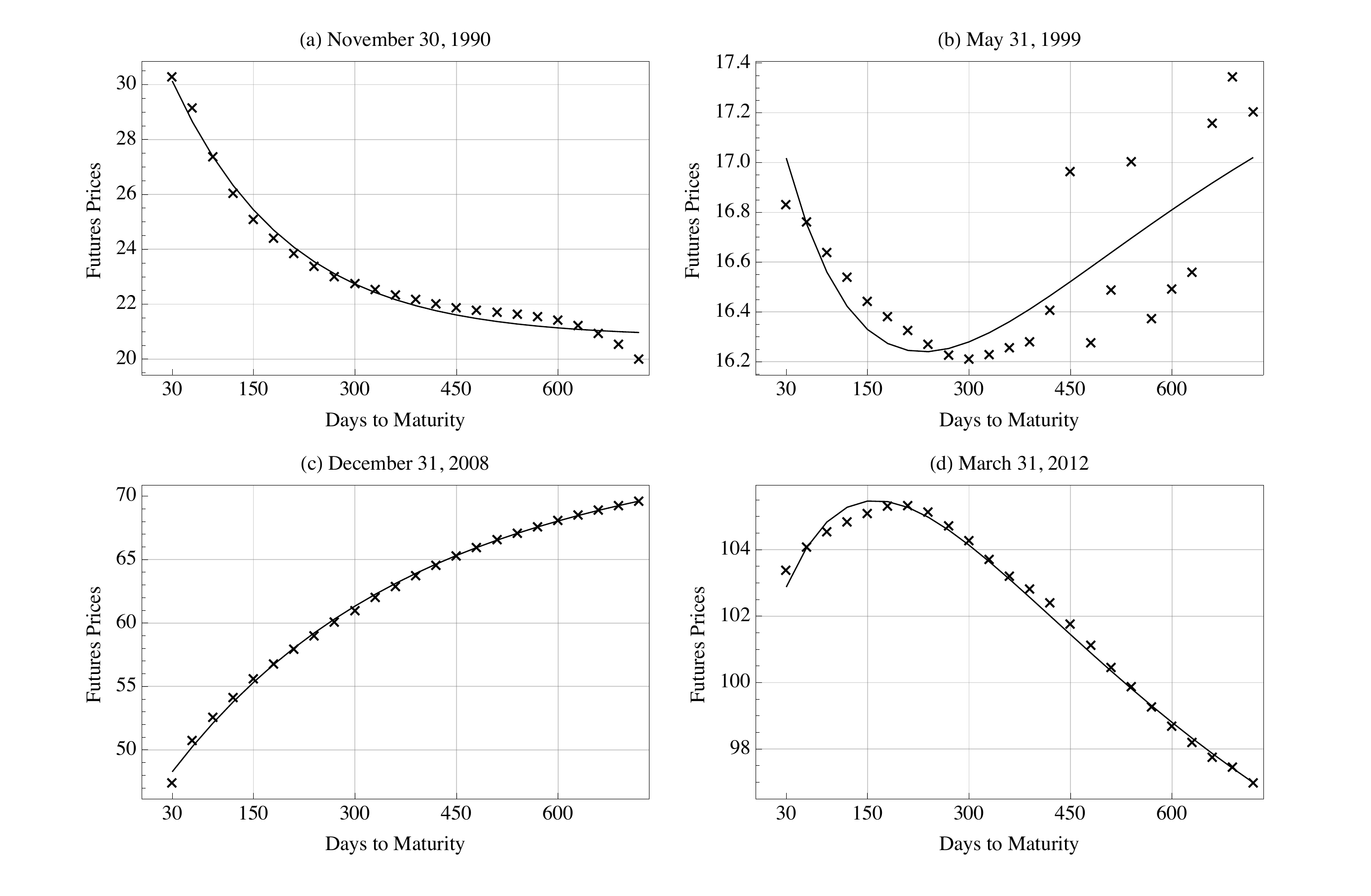}    
    \caption{Examples of term structures of futures contracts on crude oil fitted by Dynamic Nelson-Siegel model: (a) November 30, 1990, (b) May 31, 1999, (c) December 31, 2008 and (d) March 31, 2012.} 
    \label{fig:shapes} 
\end{figure} 
\cite{diebold2006} propose to forecast the factor loadings using autoregressive and vector-autoregressive models, with random walk as a benchmark. One of the directly visible features of the factor loadings is its non-stationarity. Stationarity is rejected for the level factor, and the two remaining factors are at the boarder. Whereas this makes further autoregressive analysis problematic, it is part of the motivation for the usage of neural networks, which does not need to assume stationary time series. In addition, factors may contain nonlinearities, which are not captured by simple linear time series analysis. 

Before we turn to the main part of the analysis, forecasting, we illustrate the fit of dynamic Nelson-Siegel model on the crude oil futures in Figure \ref{fig:shapes}. Term structures are generally fitted with high degree of accuracy for all curve shapes. Similarly to \cite{diebold2006}, in case of term structure with multiple local extremes (as during May 1999), the approximation is not so accurate. 

\section{Forecasting the term structure with neural networks}

To obtain the future term structure forecasts from dynamic Nelson-Siegel model, \cite{diebold2006} propose to forecast individual $\beta_t$ coefficients using linear autoregressive (AR) and Vector AR (VAR) models. In this work, we propose to forecast the individual coefficients on factor loadings using artificial neural networks. The motivation is straightforward, as $\beta_t$ coefficients are not stationary for the crude oil futures, and may further contain nonlinear dependence. Linear models are not able to capture these features well, hence we hypothesize that our proposed approach will yield more accurate forecasts. Similarly to \cite{diebold2006}, forecast of futures price with forecast horizon $h$ will be calculated as
\begin{equation}
\widehat{p}_{t+h}(\tau) = \widehat{\beta}_{0,t+h} + \widehat{\beta}_{1, t+h} \left( \frac{1-e^{-\lambda^* \tau}}{\lambda^* \tau} \right) + \widehat{\beta}_{2, t+h} \left( \frac{1-e^{-\lambda^* \tau}}{\lambda^* \tau} - e^{-\lambda^* \tau} \right),
\end{equation}
where $\widehat{\beta}_{i,t+h}$ are coefficients to be predicted. Both AR and VAR models used for prediction by \cite{diebold2006} are developed to capture linear features of the time-series. Hence using them for forecasting coefficients on factor loadings, one assumes that they are generated by a linear processes. This is not the case of Artificial neural networks (ANN), as ANNs do not require any assumptions about statistical properties of underlying series for their proper application. ANNs may be viewed as a generalization of these classical approaches, which allows us to model different type of nonlinearities in the data.

Although neural networks imitating neural processing in brain activation, are primarily associated with biological systems and successfully applied in numerous fields, such as pattern recognition, medical diagnostics, many econometricians argue that the approach is a black box. Together with the fact that one must make arbitrary decisions about the implementation of the network, i.e., the number of hidden layers, the choice of transformation functions, the number of neurons, etc., neural networks are still not commonly used for financial time series modeling, and we are pioneering their use in the term structure forecasting.

Abandoning these concerns, we use neural network as a generalized nonlinear regression, being able to describe the complex patterns in time series of curvature parameters. Like other linear or nonlinear methods, a neural network relates a set of input variables, say lags of time series, to output -- in our case the forecast. The only difference between network and other models is that the approximating function uses one or more so-called hidden layers, in which the input variables are squashed or transformed by a special function.

The most widely used artificial neural network in financial applications with one hidden layer \citep{Hornik1989359} is the feed-forward neural network. The general feed-forward or multi-layered perceptron (MLP) network we use for forecasting of $\widehat{\beta}_{t+h}$ coefficient may be described by the following equations:
\begin{eqnarray}
\widehat{\beta}_{t+h}&=&\gamma_{0}+\sum_{k=1}^{k^{*}}\gamma_{k}\Lambda(n_{k,t}) \\
\Lambda(n_{k,t})&=&\frac{1}{1+e^{-n_{k,t}}} \\
\label{fnn} 
n_{k,t}&=&\omega_{k,0}+\sum_{i=0}^{m+1}\omega_{k,i}\widehat{\beta}_{t-i}
\end{eqnarray}
with $k^{*}$ neurons $n_{k,t}$, and $\omega_{k,i}$ representing a coefficient vector or \textit{weights} vector to be found. The variable $n_{k,t}$ consisting of $m+1$ lags of time series being forecast, is squashed by the hyperbolic tangent transfer function and becomes a neuron $\Lambda(n_{k,t})$. Next, the set of $k^{*}$ neurons are combined linearly with the vector of coefficients $\{\gamma_{k}\}_{k=1}^{k^{*}}$ to form the final output, which is the forecast of the $\widehat{\beta}_{t+h}$ coefficient on factor loadings from the dynamic Nelson-Siegel model. The general feed-forward network is the workhorse of the neural network modeling approach in finance industry, as almost all researchers begin with this network as the first alternative to linear models. 

Note that AR is a simple special case within this framework if transformation $\Lambda(n_{k,t})$ is skipped (i.e. $\Lambda(n_{k,t})=n_{t,k}$) and one neuron that contains a linear approximation function is used. Therefore, in addition to classical linear models, there are neurons that process the inputs to improve the predictions. 

To be able to approximate the target function, the neural network must be able to \textit{``learn"}. The process of learning is defined as the adjustment of weights using a learning algorithm. The main goal of the learning process is to minimize the sum of the prediction errors for all training examples. The training phase is thus an unconstrained nonlinear optimization problem, where the goal is to find the optimal set of weights of the parameters by solving the minimization problem:
\begin{equation}
\min\{\Psi(\omega):\omega\in\mathbb{R}^{n}\},
\end{equation}
where $\Psi:\mathbb{R}^{n} \rightarrow \mathbb{R}^{n}$ is a continuously differentiable error function. There are several ways of minimizing $\Psi(\omega)$, but basically we are searching for the gradient $G=\nabla\Psi(\omega)$ of function $\Psi$, which is the vector of the first partial derivatives of the error function $\Psi(\omega)$ with respect to the weight vector $\omega$. Furthermore, the gradient specifies a direction that produces the steepest increase in $\Psi$. The negative of this vector thus provides us the direction of steepest decrease. 

Nevertheless, the traditional gradient descent algorithms often fail in learning intricate patterns in the data efficiently due to many possible initial settings. One of the efficient methods for learning the patterns in feed-forward neural networks, which we use, is the Levenberg-Marquardt back-propagation.

\subsection{Focused time-delay neural network}

To be able to fully explore the time dependence in time series, we use a simple extension of the feed-forward framework, as dynamic neural networks are capable to learn dynamics of time series relationships more effectively. Time-Delay Neural Network is a feed-forward network with a tapped delay line at the input. It is similar to a multilayer perceptron as all connections feed forward. In addition, the inputs to any node consist of the outputs of earlier nodes from previous time steps. This is generally implemented using tap-delay lines. 

Most straightforward general dynamic neural networks is the class, which have delays only on the input units known as Focused Time-Delay Neural Network \citep{clouse1997time}. It consists of set of feed-forward networks with tapped delay line capturing autoregressive property of inspected series. We propose to use the Focused Time-Delay Neural Network (FTDNN) for forecasting of $\beta_t$ loadings. The delay $\Delta$ is introduced to the Equation \ref{fnn} as
\begin{equation}
\label{ftdnn}
n_{k,t}=\omega_{k,0}+\sum_{i=0}^{m+1}\omega_{k,i}\widehat{\beta}_{t-(i-1)\Delta}
\end{equation}
 
In order to forecast three time series of $\beta_t$ coefficients estimated by Nelson-Siegel model, we will naturally use three separate networks. To prevent over-fitting, we use cross-validation over time with fixed window. The best model is always chosen based on the cross-validation scheme. In-sample (training and validation) and out-of-sample (testing) datasets are chosen in usual ratio of 60\%, 20\%, and 20\% for training, validation, and testing respectively. In terms of the out-of-sample forecast period, we start to forecast the futures prices from 2010. The same period is also used for forecasts from competing models defined in next section.

Input layer consists of $m$ lags relevant for forecast, where $m$ can be determined by inspecting respective sample autocorrelation function. In order to retain comparability of forecast results with AR(1) and VAR(1) models, we use one lag. A simple network with one hidden layer consisting of up to 20 hidden neurons is considered. Output neuron is $l$-th step-ahead forecast of particular $\beta_t$ coefficient: 1-month, 3-month, 6-,month and 12-month-ahead forecasts have been examined. Final decision about network structure was made according to Hannan-Quinn information criterion\footnote{$HQIC = \left[ \ln \left( \sum\limits_{t=1}^{N} \frac{(\beta_t - \hat{\beta}_t)^2}{N} \right) \right] + \frac{k(\ln(\ln(N)))}{N}$}, as it punishes networks with excess number of parameters.

\section{Out-of-sample forecasting performance}
\label{sec:six}

\subsection{Competing models}
The main interest of this work is in assessing the out-of-sample forecasting performance of neural networks in forecasting term structure of crude oil futures. Naturally, we asses the performance in relative terms to a competing models used by the literature. The first competing model we consider is a simple AR(1) process for all three $\widehat{\beta}_{i,t+h}$ coefficients $i=\{1,2,3\}$:
\begin{equation}
    \widehat{\beta}_{i,t+h} = \widehat{c}_i + \widehat{\gamma}_i \widehat{\beta}_{it},
\end{equation}
where coefficients $\widehat{c}_i$ and $\widehat{\gamma}_i$ are obtained by regressing $\widehat{\beta}_{i,t}$ on $\widehat{\beta}_{i,t-h}$ and an intercept. Factor loadings $\widehat{\beta}_{i,t+h}$ may generally contain unit root, which will result in poor forecasts due to large possible biases in estimates. Still, the model is used in the literature modeling yield curves and term structures.

A second benchmark model for forecasting term structure we consider is vector autoregressive model, where
\begin{equation}
    \widehat{\beta}_{t+h} = \widehat{c} + \widehat{\Gamma} \widehat{\beta}_{t},
\end{equation}
with $\widehat{c}$, and $\widehat{\Gamma}$ holding coefficients to be estimated. In case of autoregressive model, issues implied by potential unit root presence in one of the series are not so severe. However, unrestricted VAR models perform quite poorly in forecasting tasks. Poor performance is caused mainly by danger of over-parametrization due to large number of parameters. \cite{diebold2006} also note that factors do not share cross-correlation structure, hence we should not expect VAR(1) model to produce superior forecasts. In case of term structures, the situation is different, as the coefficients share interaction to be modeled. 

As a final, benchmark model, we consider Random Walk, where the expected forecast is the previous lag 
\begin{equation}
    \widehat{\beta}_{i,t+h} = \beta_{it}.
\end{equation}

All four models are used to forecast the term structure of crude oil futures, both in one-step-ahead and multi-step-ahead predictions (we consider 1-,3-,6-, and 12-months-ahead).

\subsection{Evaluation of forecasts}
To statistically compare the accuracy of the forecasts from different models, we employ two common loss functions, namely the root mean square error (RMSE) and the mean absolute error (MAE). The measures are calculated for the $t=1,\ldots,T$ forecasts as
\begin{equation}
RMSE = \sqrt{\frac{1}{N} \sum_{i=1}^{T}\left(\widehat{p}_{t+i} - p_{t+i} \right)^2}
\end{equation} 
\begin{equation}
MAE = \frac{1}{N} \sum_{i=1}^{T}\left|\widehat{p}_{t+i} - p_{t+i} \right|
\end{equation} 

As discussed by \cite{nomikos2011}, these metrics do not provide information about the asymmetry of the errors. While asymmetric errors are commonly found by the volatility literature, it may be also of interest to see if the models do not over-, or under-predict the term structures systematically. For example \cite{nomikos2011,wang2012,barunik2014coupling} find majority of forecasting models to over-predict the volatility on petroleum markets. The bias then translates to direct economic losses. Hence, as suggested by \cite{nomikos2011}, we employ two additional mean mixed error (MME) loss functions \citep{brailsford1996evaluation} to assess the forecasts. These functions use a mixture of positive and negative forecast errors with different weights allowing us to discover the cases if the model tends to over- or under-predict 
\begin{equation}
MME(O) = \frac{1}{N} \left( \sum_{i\in U}\left|\widehat{p}_{t+i} - p_{t+i} \right| + \sum_{i\in O}\sqrt{\left|\widehat{p}_{t+i} - p_{t+i} \right|}\right),
\end{equation} 
\begin{equation}
MME(U) = \frac{1}{N} \left( \sum_{i\in U}\sqrt{\left|\widehat{p}_{t+i} - p_{t+i} \right|} + \sum_{i\in O}\left|\widehat{p}_{t+i} - p_{t+i} \right| \right),
\end{equation} 
 where $U$ is the set containing under-predictions and $O$ is the set containing over-predictions.
 
To test significant differences of loss functions from competing models, we use the Model Confidence Set (MCS) methodology of \cite{hansen2011model}. Given a set of forecasting models, $\mathcal{M}_{0}$, we identify the model confidence set $\widehat{\mathcal{M}}^*_{1-\alpha} \subset \mathcal{M}_{0}$, which is the set of models that contain the ``best" forecasting model given a level of confidence $\alpha$. For a given model $i \in \mathcal{M}_{0}$, the $p$-value is the threshold confidence level. Model $i$ belongs to the MCS only if $\widehat{p}_i \ge \alpha$. MSC methodology repeatedly tests the null hypothesis of equal forecasting accuracy 
$$H_{0,\mathcal{M}}:E[\mathcal{L}_{i,t}-\mathcal{L}_{j,t}]=0, \hspace{1cm} \text{for all } i,j\in\mathcal{M}$$ 
with $\mathcal{L}_{i,t}$ being an appropriate loss function of the $i$-th model. Starting with the full set of models, $\mathcal{M}=\mathcal{M}_0$, this procedure sequentially eliminates the worst-performing model from $\mathcal{M}$ when the null is rejected. The surviving set of models then belong to the model confidence set $\widehat{\mathcal{M}}^*_{1-\alpha}$. Following \cite{hansen2011model}, we implement the MCS using a stationary bootstrap with an average block length of 20 days.\footnote{We have used different block lengths, including the ones depending on the forecasting horizons, to assess the robustness of the results, without any change in the final results. These results are available from the authors upon request.}



\subsection{Discussion of the results}

Four forecasting models -- focused time-delay neural network (FTDNN), AR(1), VAR(1) and random walk (RW) -- are used to forecast the term structure of crude oil futures, both in one-step-ahead and multi-step-ahead predictions. We begin with discussion of aggregate results. Average RMSE of forecasts over all maturities in Table \ref{tab:avgRMSE} reveals that FTDNN produces forecasts with the lowest errors for all forecasting horizons considered. Second-best forecasting model is AR(1) model, confirming conclusions of \cite{diebold2006} on the yield curves data who find AR(1) model to outperform both VAR(1) and RW.
 
While average results provide us with the first notion of how the models perform against each other, Tables \ref{tab:RMSE} and \ref{tab:MAE} in Appendix A provide summary of forecast performance for individual maturities. For better clarity of the results, we report RMSE and MAE relative to the respective statistics from RW as a benchmark model. A simple ratio tells us quickly, how the model under evaluation compares to the benchmark Random Walk. Moreover, the Model Confidence Set is found across all models for all time to maturities and multi-step-ahead forecasts. 

\begin{table}[t]
\footnotesize
  \centering
     \begin{tabular}{rrrrr}
    \toprule
    Horizon & FTDNN & AR(1) & VAR(1) & RW \\
    \midrule
    1 month & \textbf{4,398} & 4,708 & 4,971 & 4,772 \\
    3 months & \textbf{6,077} & 7,572 & 8,060 & 7,952 \\
    6 months & \textbf{6,425} & 8,868 & 10,362 & 10,140 \\
    12 months & \textbf{7,881} & 7,947 & 11,487 & 9,841 \\
    \bottomrule
    \end{tabular}
    \caption{Average RMSE across all constant maturities. }
  \label{tab:avgRMSE}%
\end{table}

In case of one-month-ahead forecast, FTDNN yields lowest RMSE and MAE in comparison to the rest of the models. FTDNN is the only model in the Model Confidence Set for maturities lower than 630 days according to RMSE. For longer maturities considered, specifically 660, 690, and 720, AR(1), and RW belong to the Model Confidence Set, while VAR(1) is rejected all the times. Looking at MAE, the situation is very similar with only difference, that for maturities longer than 420, all FTDNN, AR(1), and RW models produce statistically indistinguishable forecasts, while FTDNN produces the lowest average statistics.

The difference between FTDNN and all the other models is even more pronounced when forecasting 3-months-ahead, where the forecasts from FTDNN are the only forecasts which are included as best forecasts using MCS for all times to maturity. This means that FTDNN decisively produces significantly better forecasts than all other models at all maturities. 

Longer forecasts for 6-months-ahead show that FTDNN produces even larger improvements in terms of RMSE and MAE in shorter horizons, where it is the only model belonging to the MCS. The longer the horizon, the lower the gains from the FTDNN against all other models are. While FTDNN produces the lowest average RMSE and MAE, none of the models can be rejected from the Model Confidence Set for maturities larger than 300 days. This means that all models produce statistically similar 6-months-ahead forecast for longer horizons. 

The longest horizon forecasts of one year show similar results to the 6-month forecast, with VAR(1) and RW being rejected from Model Confidence Set for all maturities. For the short maturities, the FTDNN produces the best forecasts, while for longer maturities, AR(1) is included in the MCS as well. 

Summarizing the results from RMSE and MAE, we can see that FTDNN produces the forecasts with significantly lowest errors in comparison to other competing models for short maturities and short forecasting horizon. For longer maturities than 300, and longer forecasting horizon, other models play role. Often, forecasts from AR(1) model can not be statistically distinguished from the forecasts from FTDNN. We need to note here that FTDNN includes only one delayed input to make the model comparable to AR(1) and VAR(1) strategies used by the literature, and the forecasts will even improve with increasing number of lags in the FTDNN. While we have experimented with number of lags, and obtained even lower errors, the sample size of the data does not allow us to rigorously study these models, and we leave it for future research.

To see if the models do not over-, or under-predict the term structures, we employ the MME(U) and MME(O) statistics.

Table \ref{tab:MMEU} shows the average number of cases when the error from the model is negative for all models across forecasting horizons, and maturities. Table \ref{tab:MMEO} shows the average number of cases when the error from the model is positive. In addition, asymmetric errors are tested using MME(U), and MME(O) in the MCS framework. In short, Table \ref{tab:MMEU} shows if the models tends to under-predict the term structures, while Table \ref{tab:MMEO} tends to over-predict the term structures. 

The important observation form the asymmetric loss functions is that models in general produce symmetric forecasts in the short term forecasts, and short times to maturities. With longer time to maturities, the FTDNN tends to under-predict at 1-month and 12-month-ahead forecasts, while over-predict at 3-month, and 6-month-ahead forecasts. AR(1) tends to generally under-predict at all forecasting horizons. For the longest forecasting horizon of 1 year, and longer time to maturities, AR(1) largely over-predicts the futures prices. The results reveal the similar pattern in terms of forecast comparisons. FTDNN is never rejected from the Model Confidence Set.

\section{Conclusion}
\label{conclusion}

This paper investigates the properties of crude oil markets term structure, and propose dynamic neural networks for their forecasting.
 
The term structure of crude oil futures prices exhibits very similar behavior to government bonds yield curve, and three-factor dynamic Nelson-Siegel model \citep{diebold2006} used by the literature for modeling yield curves captures the shapes of the term structure very well. We further forecast the factors using dynamic neural network. 

Proposed framework yields significant improvements in the futures prices forecasts when compared to other benchmark models. We show the performance on the 1-month, 3-month, 6-month and 12-month forecasting horizons. Forecasting errors from our approach have moreover traceable patterns. For fixed forecasting horizon, the deviation between forecast and observed futures price decreases as time to maturity increases. Furthermore, for more distant forecast horizons the deviation on average expectedly increases.
 
In summary, this work has shown that crude oil term structure can be successfully modeled and predicted by parsimonious Nelson-Siegel model primarily developed for interest rates coupled with generalized regression framework of neural networks. The future research will show if our results hold for other commodities as well. An interesting and important approach would also be to use the framework to study the commonalities between factors across various commodities.  

\bibliographystyle{chicago}
\bibliography{NNtermstructure}

\section*{Appendix: Figures and Tables}

\begin{landscape}
\begin{table}
  \centering
    \resizebox{1.7 \textwidth}{!}{\begin{tabular}{rrrrrrrrrrrrrrrrrrrrrrrrr}
    \toprule
    \textbf{RMSE} & \multicolumn{24}{c}{Time to maturity} \\
    \midrule
          & 30    & 60    & 90    & 120   & 150   & 180   & 210   & 240   & 270   & 300   & 330   & 360   & 390   & 420   & 450   & 480   & 510   & 540   & 570   & 600   & 630   & 660   & 690   & 720 \\
\textbf{FTDNN} \\ 
1 M & \textbf{ 0.81}& \textbf{ 0.81}& \textbf{ 0.80}& \textbf{ 0.80}& \textbf{ 0.80}& \textbf{ 0.79}& \textbf{ 0.79}& \textbf{ 0.79}& \textbf{ 0.78}& \textbf{ 0.78}& \textbf{ 0.77}& \textbf{ 0.77}& \textbf{ 0.77}& \textbf{ 0.77}& \textbf{ 0.77}& \textbf{ 0.77}& \textbf{ 0.77}& \textbf{ 0.77}& \textbf{ 0.77}& \textbf{ 0.77}& \textbf{ 0.78}& \textbf{ 0.79}& \textbf{ 0.80}& \textbf{ 0.83} \\ 
3 M & \textbf{ 0.73}& \textbf{ 0.73}& \textbf{ 0.73}& \textbf{ 0.72}& \textbf{ 0.72}& \textbf{ 0.72}& \textbf{ 0.72}& \textbf{ 0.72}& \textbf{ 0.72}& \textbf{ 0.72}& \textbf{ 0.72}& \textbf{ 0.72}& \textbf{ 0.72}& \textbf{ 0.73}& \textbf{ 0.73}& \textbf{ 0.73}& \textbf{ 0.74}& \textbf{ 0.74}& \textbf{ 0.74}& \textbf{ 0.75}& \textbf{ 0.75}& \textbf{ 0.76}& \textbf{ 0.77}& \textbf{ 0.78} \\ 
6 M & \textbf{ 0.59}& \textbf{ 0.59}& \textbf{ 0.58}& \textbf{ 0.58}& \textbf{ 0.58}& \textbf{ 0.58}& \textbf{ 0.58}& \textbf{ 0.59}& \textbf{ 0.59}& \textbf{ 0.60}& \textbf{ 0.61}& \textbf{ 0.61}& \textbf{ 0.62}& \textbf{ 0.62}& \textbf{ 0.63}& \textbf{ 0.64}& \textbf{ 0.64}& \textbf{ 0.65}& \textbf{ 0.66}& \textbf{ 0.66}& \textbf{ 0.67}& \textbf{ 0.67}& \textbf{ 0.68}& \textbf{ 0.69} \\ 
12 M & \textbf{ 0.74}& \textbf{ 0.75}& \textbf{ 0.75}& \textbf{ 0.75}& \textbf{ 0.75}& \textbf{ 0.75}& \textbf{ 0.75}& \textbf{ 0.75}& \textbf{ 0.75}& \textbf{ 0.75}& \textbf{ 0.75}& \textbf{ 0.75}& \textbf{ 0.75}& \textbf{ 0.75}& \textbf{ 0.75}& \textbf{ 0.75}& \textbf{ 0.74}& \textbf{ 0.74}& \textbf{ 0.74}& \textbf{ 0.74}& \textbf{ 0.75}& \textbf{ 0.75}& \textbf{ 0.75}& \textbf{ 0.75} \\ 
\textbf{AR(1)} \\ 
1 M &  0.98 &  0.98 &  0.99 &  0.99 &  0.99 &  0.99 &  0.99 &  0.99 &  0.99 &  0.99 &  0.98 &  0.98 &  0.98 &  0.98 &  0.98 &  0.99 &  0.99 &  0.99 &  0.99 &  0.99 &  0.99 & \textbf{ 0.99}& \textbf{ 1.00}& \textbf{ 1.02} \\ 
3 M &  0.93 &  0.94 &  0.95 &  0.95 &  0.95 &  0.95 &  0.94 &  0.94 &  0.95 &  0.95 &  0.95 &  0.95 &  0.95 &  0.95 &  0.95 &  0.96 &  0.96 &  0.96 &  0.97 &  0.96 &  0.97 &  0.97 &  0.97 &  0.98 \\ 
6 M &  0.87 &  0.87 &  0.87 &  0.87 &  0.87 &  0.86 &  0.86 &  0.86 &  0.86 & \textbf{ 0.87}& \textbf{ 0.87}& \textbf{ 0.87}& \textbf{ 0.87}& \textbf{ 0.87}& \textbf{ 0.87}& \textbf{ 0.87}& \textbf{ 0.88}& \textbf{ 0.88}& \textbf{ 0.89}& \textbf{ 0.89}& \textbf{ 0.89}& \textbf{ 0.90}& \textbf{ 0.90}& \textbf{ 0.90} \\ 
12 M &  0.87 &  0.87 &  0.85 &  0.84 & \textbf{ 0.83}& \textbf{ 0.81}& \textbf{ 0.80}& \textbf{ 0.79}& \textbf{ 0.78}& \textbf{ 0.78}& \textbf{ 0.77}& \textbf{ 0.77}& \textbf{ 0.77}& \textbf{ 0.77}& \textbf{ 0.78}& \textbf{ 0.78}& \textbf{ 0.78}& \textbf{ 0.79}& \textbf{ 0.80}& \textbf{ 0.81}& \textbf{ 0.82}& \textbf{ 0.82}& \textbf{ 0.83}& \textbf{ 0.83} \\ 
\textbf{VAR(1)} \\ 
1 M &  1.01 &  1.01 &  1.02 &  1.02 &  1.02 &  1.02 &  1.03 &  1.03 &  1.03 &  1.03 &  1.03 &  1.04 &  1.04 &  1.04 &  1.05 &  1.05 &  1.06 &  1.07 &  1.07 &  1.07 &  1.08 &  1.08 &  1.09 &  1.12 \\ 
3 M &  0.99 &  1.00 &  1.00 &  1.00 &  1.00 &  0.99 &  0.99 &  0.99 &  1.00 &  1.00 &  1.00 &  1.00 &  1.01 &  1.01 &  1.02 &  1.02 &  1.03 &  1.04 &  1.04 &  1.05 &  1.05 &  1.05 &  1.06 &  1.08 \\ 
6 M &  1.08 &  1.07 &  1.06 &  1.05 &  1.03 &  1.02 &  1.02 &  1.01 &  1.01 & \textbf{ 1.01}& \textbf{ 1.00}& \textbf{ 1.00}& \textbf{ 1.00}& \textbf{ 1.00}& \textbf{ 1.00}& \textbf{ 1.00}& \textbf{ 1.00}& \textbf{ 1.01}& \textbf{ 1.01}& \textbf{ 1.01}& \textbf{ 1.01}& \textbf{ 1.02}& \textbf{ 1.02}& \textbf{ 1.03} \\ 
12 M &  1.21 &  1.20 &  1.19 &  1.18 &  1.16 &  1.15 &  1.14 &  1.13 &  1.13 &  1.13 &  1.13 &  1.13 &  1.13 &  1.14 &  1.14 &  1.15 &  1.16 &  1.17 &  1.18 &  1.20 &  1.21 &  1.22 &  1.22 &  1.23 \\ 
\textbf{RW} \\ 
1 M &  1.00 &  1.00 &  1.00 &  1.00 &  1.00 &  1.00 &  1.00 &  1.00 &  1.00 &  1.00 &  1.00 &  1.00 &  1.00 &  1.00 &  1.00 &  1.00 &  1.00 &  1.00 &  1.00 &  1.00 &  1.00 & \textbf{ 1.00}& \textbf{ 1.00}& \textbf{ 1.00} \\ 
3 M &  1.00 &  1.00 &  1.00 &  1.00 &  1.00 &  1.00 &  1.00 &  1.00 &  1.00 &  1.00 &  1.00 &  1.00 &  1.00 &  1.00 &  1.00 &  1.00 &  1.00 &  1.00 &  1.00 &  1.00 &  1.00 &  1.00 &  1.00 &  1.00 \\ 
6 M &  1.00 &  1.00 &  1.00 &  1.00 &  1.00 &  1.00 &  1.00 &  1.00 &  1.00 & \textbf{ 1.00}& \textbf{ 1.00}& \textbf{ 1.00}& \textbf{ 1.00}& \textbf{ 1.00}& \textbf{ 1.00}& \textbf{ 1.00}& \textbf{ 1.00}& \textbf{ 1.00}& \textbf{ 1.00}& \textbf{ 1.00}& \textbf{ 1.00}& \textbf{ 1.00}& \textbf{ 1.00}& \textbf{ 1.00} \\ 
12 M &  1.00 &  1.00 &  1.00 &  1.00 &  1.00 &  1.00 &  1.00 &  1.00 &  1.00 &  1.00 &  1.00 &  1.00 &  1.00 &  1.00 &  1.00 &  1.00 &  1.00 &  1.00 &  1.00 &  1.00 &  1.00 &  1.00 &  1.00 &  1.00 \\ 
 
\bottomrule
    \end{tabular}}
    \caption{The table reports RMSE from model forecasts relative to the RMSE of RW model. The competing model has lower error than RW in case the ratio is lower than one, and vice versa. We consider multiple-step-ahead forecasts of 1,3,6, and 12 months for individual maturities using four competing models: FTDNN, AR(1), VAR(1), and RW. The Model Confidence Set (MSC) is used to compare the errors across the four competing models. The ratio is bold in case of the model belonging to the $\widehat{\mathcal{M}}^*_{10\%}$.}
  \label{tab:RMSE}
\end{table}
\end{landscape}

\begin{landscape}
\begin{table}
  \centering
    \resizebox{1.7 \textwidth}{!}{\begin{tabular}{rrrrrrrrrrrrrrrrrrrrrrrrr}

    \toprule
    \textbf{MAE} & \multicolumn{24}{c}{Time to maturity} \\
    \midrule
          & 30    & 60    & 90    & 120   & 150   & 180   & 210   & 240   & 270   & 300   & 330   & 360   & 390   & 420   & 450   & 480   & 510   & 540   & 570   & 600   & 630   & 660   & 690   & 720 \\
\textbf{FTDNN} \\ 
1 M & \textbf{ 0.79}& \textbf{ 0.80}& \textbf{ 0.81}& \textbf{ 0.81}& \textbf{ 0.83}& \textbf{ 0.83}& \textbf{ 0.83}& \textbf{ 0.83}& \textbf{ 0.83}& \textbf{ 0.84}& \textbf{ 0.83}& \textbf{ 0.83}& \textbf{ 0.83}& \textbf{ 0.83}& \textbf{ 0.84}& \textbf{ 0.84}& \textbf{ 0.84}& \textbf{ 0.83}& \textbf{ 0.84}& \textbf{ 0.83}& \textbf{ 0.84}& \textbf{ 0.85}& \textbf{ 0.85}& \textbf{ 0.88} \\ 
3 M & \textbf{ 0.73}& \textbf{ 0.71}& \textbf{ 0.70}& \textbf{ 0.69}& \textbf{ 0.68}& \textbf{ 0.68}& \textbf{ 0.68}& \textbf{ 0.68}& \textbf{ 0.69}& \textbf{ 0.68}& \textbf{ 0.69}& \textbf{ 0.68}& \textbf{ 0.68}& \textbf{ 0.68}& \textbf{ 0.68}& \textbf{ 0.70}& \textbf{ 0.71}& \textbf{ 0.72}& \textbf{ 0.72}& \textbf{ 0.73}& \textbf{ 0.74}& \textbf{ 0.75}& \textbf{ 0.77}& \textbf{ 0.77} \\ 
6 M & \textbf{ 0.61}& \textbf{ 0.62}& \textbf{ 0.63}& \textbf{ 0.63}& \textbf{ 0.63}& \textbf{ 0.64}& \textbf{ 0.65}& \textbf{ 0.65}& \textbf{ 0.66}& \textbf{ 0.66}& \textbf{ 0.67}& \textbf{ 0.67}& \textbf{ 0.67}& \textbf{ 0.68}& \textbf{ 0.68}& \textbf{ 0.69}& \textbf{ 0.69}& \textbf{ 0.70}& \textbf{ 0.70}& \textbf{ 0.70}& \textbf{ 0.70}& \textbf{ 0.71}& \textbf{ 0.72}& \textbf{ 0.74} \\ 
12 M & \textbf{ 0.68}& \textbf{ 0.69}& \textbf{ 0.70}& \textbf{ 0.71}& \textbf{ 0.72}& \textbf{ 0.72}& \textbf{ 0.72}& \textbf{ 0.72}& \textbf{ 0.71}& \textbf{ 0.71}& \textbf{ 0.70}& \textbf{ 0.70}& \textbf{ 0.69}& \textbf{ 0.69}& \textbf{ 0.68}& \textbf{ 0.68}& \textbf{ 0.67}& \textbf{ 0.67}& \textbf{ 0.66}& \textbf{ 0.66}& \textbf{ 0.66}& \textbf{ 0.65}& \textbf{ 0.65}& \textbf{ 0.64} \\ 
\textbf{AR(1)} \\ 
1 M &  0.95 &  0.96 &  0.97 &  0.98 &  0.99 &  0.99 &  0.99 &  1.00 &  1.00 &  1.01 &  0.99 &  1.00 &  0.99 & \textbf{ 1.00}& \textbf{ 1.01}& \textbf{ 1.01}& \textbf{ 1.01}& \textbf{ 0.99}& \textbf{ 0.99}& \textbf{ 0.99}& \textbf{ 0.99}& \textbf{ 0.99}& \textbf{ 0.98}& \textbf{ 1.02} \\ 
3 M &  0.98 &  0.99 &  0.99 &  0.99 &  0.98 &  0.98 &  0.98 &  0.98 &  0.97 &  0.97 &  0.97 &  0.97 &  0.96 &  0.96 &  0.96 &  0.96 &  0.96 &  0.95 &  0.96 &  0.96 &  0.96 &  0.96 &  0.96 &  0.97 \\ 
6 M &  0.92 &  0.93 &  0.94 &  0.95 &  0.95 &  0.94 &  0.93 &  0.92 & \textbf{ 0.92}& \textbf{ 0.92}& \textbf{ 0.92}& \textbf{ 0.91}& \textbf{ 0.91}& \textbf{ 0.91}& \textbf{ 0.90}& \textbf{ 0.90}& \textbf{ 0.89}& \textbf{ 0.89}& \textbf{ 0.89}& \textbf{ 0.88}& \textbf{ 0.88}& \textbf{ 0.87}& \textbf{ 0.87}& \textbf{ 0.88} \\ 
12 M &  0.83 &  0.84 &  0.85 &  0.85 &  0.84 &  0.84 & \textbf{ 0.82}& \textbf{ 0.81}& \textbf{ 0.78}& \textbf{ 0.77}& \textbf{ 0.74}& \textbf{ 0.73}& \textbf{ 0.73}& \textbf{ 0.72}& \textbf{ 0.72}& \textbf{ 0.71}& \textbf{ 0.71}& \textbf{ 0.71}& \textbf{ 0.72}& \textbf{ 0.73}& \textbf{ 0.75}& \textbf{ 0.76}& \textbf{ 0.77}& \textbf{ 0.77} \\ 
\textbf{VAR(1)} \\ 
1 M &  1.02 &  1.02 &  1.03 &  1.03 &  1.04 &  1.04 &  1.05 &  1.05 &  1.06 &  1.07 &  1.07 &  1.08 &  1.09 &  1.09 &  1.11 &  1.12 &  1.12 &  1.12 &  1.12 &  1.12 &  1.13 &  1.14 &  1.14 &  1.17 \\ 
3 M &  1.02 &  1.03 &  1.03 &  1.02 &  1.02 &  1.02 &  1.02 &  1.02 &  1.02 &  1.02 &  1.01 &  1.02 &  1.02 &  1.02 &  1.03 &  1.04 &  1.04 &  1.05 &  1.05 &  1.06 &  1.07 &  1.08 &  1.09 &  1.12 \\ 
6 M &  1.14 &  1.14 &  1.13 &  1.12 &  1.10 &  1.07 &  1.05 &  1.03 & \textbf{ 1.02}& \textbf{ 1.00}& \textbf{ 0.98}& \textbf{ 0.97}& \textbf{ 0.96}& \textbf{ 0.96}& \textbf{ 0.96}& \textbf{ 0.97}& \textbf{ 0.97}& \textbf{ 0.98}& \textbf{ 0.98}& \textbf{ 0.98}& \textbf{ 0.99}& \textbf{ 1.00}& \textbf{ 1.01}& \textbf{ 1.03} \\ 
12 M &  1.17 &  1.16 &  1.16 &  1.16 &  1.16 &  1.16 &  1.15 &  1.14 &  1.14 &  1.14 &  1.14 &  1.14 &  1.16 &  1.17 &  1.19 &  1.20 &  1.21 &  1.22 &  1.25 &  1.26 &  1.27 &  1.28 &  1.28 &  1.28 \\ 
\textbf{RW} \\ 
1 M &  1.00 &  1.00 &  1.00 &  1.00 &  1.00 &  1.00 &  1.00 &  1.00 &  1.00 &  1.00 &  1.00 &  1.00 &  1.00 & \textbf{ 1.00}& \textbf{ 1.00}& \textbf{ 1.00}& \textbf{ 1.00}& \textbf{ 1.00}& \textbf{ 1.00}& \textbf{ 1.00}& \textbf{ 1.00}& \textbf{ 1.00}& \textbf{ 1.00}& \textbf{ 1.00} \\ 
3 M &  1.00 &  1.00 &  1.00 &  1.00 &  1.00 &  1.00 &  1.00 &  1.00 &  1.00 &  1.00 &  1.00 &  1.00 &  1.00 &  1.00 &  1.00 &  1.00 &  1.00 &  1.00 &  1.00 &  1.00 &  1.00 &  1.00 &  1.00 &  1.00 \\ 
6 M &  1.00 &  1.00 &  1.00 &  1.00 &  1.00 &  1.00 &  1.00 &  1.00 & \textbf{ 1.00}& \textbf{ 1.00}& \textbf{ 1.00}& \textbf{ 1.00}& \textbf{ 1.00}& \textbf{ 1.00}& \textbf{ 1.00}& \textbf{ 1.00}& \textbf{ 1.00}& \textbf{ 1.00}& \textbf{ 1.00}& \textbf{ 1.00}& \textbf{ 1.00}& \textbf{ 1.00}& \textbf{ 1.00}& \textbf{ 1.00} \\ 
12 M &  1.00 &  1.00 &  1.00 &  1.00 &  1.00 &  1.00 &  1.00 &  1.00 &  1.00 &  1.00 &  1.00 &  1.00 &  1.00 &  1.00 &  1.00 &  1.00 &  1.00 &  1.00 &  1.00 &  1.00 &  1.00 &  1.00 &  1.00 &  1.00 \\ 

\bottomrule
    \end{tabular}}
    \caption{The table reports MAE from model forecasts relative to the MAE of RW model. The competing model has lower error than RW in case the ratio is lower than one, and vice versa. We consider multiple-step-ahead forecasts of 1,3,6, and 12 months for individual maturities using four competing models: FTDNN, AR(1), VAR(1), and RW. The Model Confidence Set (MSC) is used to compare the errors across the four competing models. The ratio is bold in case of the model belonging to the $\widehat{\mathcal{M}}^*_{10\%}$.}
  \label{tab:MAE}
\end{table}
\end{landscape}

\begin{landscape}
\begin{table}
  \centering
    \resizebox{1.7 \textwidth}{!}{\begin{tabular}{rrrrrrrrrrrrrrrrrrrrrrrrr}

    \toprule
    \textbf{MME(U)} & \multicolumn{24}{c}{Time to maturity} \\
    \midrule
          & 30    & 60    & 90    & 120   & 150   & 180   & 210   & 240   & 270   & 300   & 330   & 360   & 390   & 420   & 450   & 480   & 510   & 540   & 570   & 600   & 630   & 660   & 690   & 720 \\
\textbf{FTDNN} \\ 
1 M & \textbf{ 0.53}& \textbf{ 0.56}& \textbf{ 0.60}& \textbf{ 0.61}& \textbf{ 0.61}& \textbf{ 0.58}& \textbf{ 0.58}& \textbf{ 0.58}& \textbf{ 0.56}& \textbf{ 0.60}& \textbf{ 0.56}& \textbf{ 0.56}& \textbf{ 0.58}& \textbf{ 0.60}& \textbf{ 0.60}& \textbf{ 0.60}& \textbf{ 0.58}& \textbf{ 0.54}& \textbf{ 0.58}& \textbf{ 0.56}& \textbf{ 0.56}& \textbf{ 0.58}& \textbf{ 0.54}& \textbf{ 0.54} \\ 
3 M & \textbf{ 0.55}& \textbf{ 0.53}& \textbf{ 0.55}& \textbf{ 0.56}& \textbf{ 0.56}& \textbf{ 0.56}& \textbf{ 0.53}& \textbf{ 0.53}& \textbf{ 0.51}& \textbf{ 0.51}& \textbf{ 0.51}& \textbf{ 0.49}& \textbf{ 0.49}& \textbf{ 0.47}& \textbf{ 0.47}& \textbf{ 0.45}& \textbf{ 0.44}& \textbf{ 0.44}& \textbf{ 0.42}& \textbf{ 0.40}& \textbf{ 0.40}& \textbf{ 0.40}& \textbf{ 0.40}& \textbf{ 0.38} \\ 
6 M & \textbf{ 0.40}& \textbf{ 0.44}& \textbf{ 0.42}& \textbf{ 0.44}& \textbf{ 0.46}& \textbf{ 0.46}& \textbf{ 0.46}& \textbf{ 0.42}& \textbf{ 0.40}& \textbf{ 0.40}& \textbf{ 0.40}& \textbf{ 0.40}& \textbf{ 0.40}& \textbf{ 0.40}& \textbf{ 0.38}& \textbf{ 0.37}& \textbf{ 0.37}& \textbf{ 0.37}& \textbf{ 0.37}& \textbf{ 0.33}& \textbf{ 0.33}& \textbf{ 0.33}& \textbf{ 0.31}& \textbf{ 0.31} \\ 
12 M & \textbf{ 0.74}& \textbf{ 0.72}& \textbf{ 0.72}& \textbf{ 0.70}& \textbf{ 0.72}& \textbf{ 0.74}& \textbf{ 0.72}& \textbf{ 0.74}& \textbf{ 0.72}& \textbf{ 0.72}& \textbf{ 0.67}& \textbf{ 0.63}& \textbf{ 0.61}& \textbf{ 0.61}& \textbf{ 0.61}& \textbf{ 0.59}& \textbf{ 0.59}& \textbf{ 0.61}& \textbf{ 0.61}& \textbf{ 0.61}& \textbf{ 0.61}& \textbf{ 0.61}& \textbf{ 0.61}& \textbf{ 0.59} \\ 
\textbf{AR(1)} \\ 
1 M &  0.51 &  0.51 &  0.53 &  0.53 &  0.54 &  0.54 &  0.54 &  0.56 &  0.56 &  0.61 &  0.60 &  0.60 & \textbf{ 0.60}& \textbf{ 0.63}& \textbf{ 0.61}& \textbf{ 0.63}& \textbf{ 0.61}& \textbf{ 0.60}& \textbf{ 0.61}& \textbf{ 0.60}& \textbf{ 0.60}& \textbf{ 0.60}& \textbf{ 0.58}& \textbf{ 0.60} \\ 
3 M & \textbf{ 0.60}& \textbf{ 0.60}& \textbf{ 0.60}& \textbf{ 0.62}& \textbf{ 0.62}& \textbf{ 0.64}&  0.62 & \textbf{ 0.60}& \textbf{ 0.60}&  0.60 & \textbf{ 0.60}& \textbf{ 0.58}& \textbf{ 0.60}& \textbf{ 0.58}& \textbf{ 0.58}& \textbf{ 0.58}& \textbf{ 0.58}& \textbf{ 0.55}& \textbf{ 0.53}& \textbf{ 0.53}& \textbf{ 0.53}& \textbf{ 0.53}& \textbf{ 0.53}& \textbf{ 0.53} \\ 
6 M & \textbf{ 0.69}& \textbf{ 0.67}& \textbf{ 0.65}& \textbf{ 0.67}& \textbf{ 0.67}& \textbf{ 0.67}& \textbf{ 0.67}& \textbf{ 0.60}& \textbf{ 0.60}& \textbf{ 0.60}& \textbf{ 0.60}& \textbf{ 0.60}& \textbf{ 0.60}& \textbf{ 0.60}& \textbf{ 0.60}& \textbf{ 0.60}& \textbf{ 0.60}& \textbf{ 0.60}& \textbf{ 0.58}& \textbf{ 0.58}& \textbf{ 0.58}& \textbf{ 0.56}& \textbf{ 0.56}& \textbf{ 0.54} \\ 
12 M &  0.83 &  0.83 &  0.83 &  0.83 & \textbf{ 0.78}& \textbf{ 0.78}& \textbf{ 0.78}& \textbf{ 0.78}& \textbf{ 0.78}& \textbf{ 0.74}& \textbf{ 0.72}& \textbf{ 0.65}& \textbf{ 0.61}& \textbf{ 0.59}& \textbf{ 0.59}& \textbf{ 0.59}& \textbf{ 0.52}& \textbf{ 0.48}& \textbf{ 0.50}& \textbf{ 0.43}& \textbf{ 0.39}& \textbf{ 0.37}& \textbf{ 0.33}& \textbf{ 0.33} \\ 
\textbf{VAR(1)} \\ 
1 M &  0.53 &  0.53 &  0.54 &  0.54 &  0.54 &  0.53 &  0.54 &  0.54 &  0.51 &  0.47 &  0.47 &  0.47 &  0.47 &  0.47 &  0.47 &  0.46 &  0.46 &  0.46 &  0.44 &  0.44 &  0.44 &  0.44 &  0.44 &  0.44 \\ 
3 M & \textbf{ 0.62}& \textbf{ 0.65}& \textbf{ 0.65}& \textbf{ 0.60}& \textbf{ 0.58}& \textbf{ 0.58}&  0.58 & \textbf{ 0.58}& \textbf{ 0.56}&  0.55 & \textbf{ 0.55}&  0.53 &  0.49 &  0.49 &  0.47 &  0.45 &  0.45 &  0.44 &  0.44 &  0.42 &  0.42 &  0.38 &  0.36 &  0.38 \\ 
6 M & \textbf{ 0.75}& \textbf{ 0.75}& \textbf{ 0.75}& \textbf{ 0.75}& \textbf{ 0.73}& \textbf{ 0.73}& \textbf{ 0.69}& \textbf{ 0.67}& \textbf{ 0.67}& \textbf{ 0.67}& \textbf{ 0.63}& \textbf{ 0.60}& \textbf{ 0.60}& \textbf{ 0.56}& \textbf{ 0.52}& \textbf{ 0.54}& \textbf{ 0.56}& \textbf{ 0.54}& \textbf{ 0.52}& \textbf{ 0.50}& \textbf{ 0.48}& \textbf{ 0.46}&  0.46 &  0.42 \\ 
12 M &  0.78 &  0.78 &  0.76 &  0.76 &  0.74 &  0.72 &  0.72 &  0.65 &  0.61 &  0.59 &  0.57 &  0.50 &  0.50 &  0.48 &  0.46 &  0.46 &  0.43 &  0.43 &  0.43 &  0.41 &  0.39 &  0.35 &  0.35 &  0.35 \\ 
\textbf{RW} \\ 
1 M &  0.53 &  0.54 &  0.54 &  0.51 &  0.51 &  0.51 &  0.53 &  0.53 &  0.53 &  0.51 &  0.53 &  0.53 & \textbf{ 0.51}& \textbf{ 0.51}& \textbf{ 0.47}& \textbf{ 0.47}& \textbf{ 0.51}& \textbf{ 0.49}& \textbf{ 0.51}& \textbf{ 0.51}& \textbf{ 0.51}& \textbf{ 0.49}& \textbf{ 0.49}& \textbf{ 0.47} \\ 
3 M &  0.53 &  0.51 &  0.51 &  0.53 &  0.53 &  0.53 &  0.53 & \textbf{ 0.51}& \textbf{ 0.51}&  0.51 & \textbf{ 0.51}&  0.49 & \textbf{ 0.51}&  0.55 &  0.55 & \textbf{ 0.53}& \textbf{ 0.51}& \textbf{ 0.53}& \textbf{ 0.53}& \textbf{ 0.51}& \textbf{ 0.51}& \textbf{ 0.51}& \textbf{ 0.51}& \textbf{ 0.51} \\ 
6 M &  0.60 &  0.60 & \textbf{ 0.58}& \textbf{ 0.54}& \textbf{ 0.50}&  0.50 &  0.50 &  0.48 &  0.48 &  0.48 &  0.48 &  0.48 &  0.46 &  0.46 &  0.42 &  0.42 &  0.42 &  0.42 &  0.42 &  0.40 &  0.40 &  0.40 &  0.40 &  0.40 \\ 
12 M &  0.59 &  0.59 &  0.59 &  0.59 &  0.57 &  0.57 &  0.50 &  0.50 &  0.46 &  0.43 &  0.41 &  0.37 &  0.37 &  0.39 &  0.39 &  0.39 &  0.37 &  0.37 &  0.37 &  0.37 &  0.33 &  0.30 &  0.30 &  0.30 \\

\bottomrule
    \end{tabular}}
    \caption{The table reports average number of cases when the error from forecasts of four competing models FTDNN, AR(1), VAR(1), and RW are negative. The higher the number from 0.5, the more under-predictions model yields. We consider multiple-step-ahead forecasts of 1,3,6, and 12 months for individual maturities. The Model Confidence Set (MSC) is used to compare the MME(U) errors across the four competing models. The ratio is bold in case of the model belonging to the $\widehat{\mathcal{M}}^*_{10\%}$.}
  \label{tab:MMEU}
\end{table}
\end{landscape}

\begin{landscape}
\begin{table}
  \centering
    \resizebox{1.7 \textwidth}{!}{\begin{tabular}{rrrrrrrrrrrrrrrrrrrrrrrrr}

    \toprule
    \textbf{MME(O)} & \multicolumn{24}{c}{Time to maturity} \\
    \midrule
          & 30    & 60    & 90    & 120   & 150   & 180   & 210   & 240   & 270   & 300   & 330   & 360   & 390   & 420   & 450   & 480   & 510   & 540   & 570   & 600   & 630   & 660   & 690   & 720 \\
\textbf{FTDNN} \\ 
1 M & \textbf{ 0.47}& \textbf{ 0.44}& \textbf{ 0.40}& \textbf{ 0.39}& \textbf{ 0.39}& \textbf{ 0.42}& \textbf{ 0.42}& \textbf{ 0.42}& \textbf{ 0.44}& \textbf{ 0.40}& \textbf{ 0.44}& \textbf{ 0.44}& \textbf{ 0.42}& \textbf{ 0.40}& \textbf{ 0.40}& \textbf{ 0.40}& \textbf{ 0.42}& \textbf{ 0.46}& \textbf{ 0.42}& \textbf{ 0.44}& \textbf{ 0.44}& \textbf{ 0.42}& \textbf{ 0.46}& \textbf{ 0.46} \\ 
3 M & \textbf{ 0.45}& \textbf{ 0.47}& \textbf{ 0.45}& \textbf{ 0.44}& \textbf{ 0.44}& \textbf{ 0.44}& \textbf{ 0.47}& \textbf{ 0.47}& \textbf{ 0.49}& \textbf{ 0.49}& \textbf{ 0.49}& \textbf{ 0.51}& \textbf{ 0.51}& \textbf{ 0.53}& \textbf{ 0.53}& \textbf{ 0.55}& \textbf{ 0.56}& \textbf{ 0.56}& \textbf{ 0.58}& \textbf{ 0.60}& \textbf{ 0.60}& \textbf{ 0.60}& \textbf{ 0.60}& \textbf{ 0.62} \\ 
6 M & \textbf{ 0.60}& \textbf{ 0.56}& \textbf{ 0.58}& \textbf{ 0.56}& \textbf{ 0.54}& \textbf{ 0.54}& \textbf{ 0.54}& \textbf{ 0.58}& \textbf{ 0.60}& \textbf{ 0.60}& \textbf{ 0.60}& \textbf{ 0.60}& \textbf{ 0.60}& \textbf{ 0.60}& \textbf{ 0.62}& \textbf{ 0.63}& \textbf{ 0.63}& \textbf{ 0.63}& \textbf{ 0.63}& \textbf{ 0.67}& \textbf{ 0.67}& \textbf{ 0.67}& \textbf{ 0.69}& \textbf{ 0.69} \\ 
12 M & \textbf{ 0.26}& \textbf{ 0.28}& \textbf{ 0.28}& \textbf{ 0.30}& \textbf{ 0.28}& \textbf{ 0.26}& \textbf{ 0.28}& \textbf{ 0.26}& \textbf{ 0.28}& \textbf{ 0.28}& \textbf{ 0.33}& \textbf{ 0.37}& \textbf{ 0.39}& \textbf{ 0.39}& \textbf{ 0.39}& \textbf{ 0.41}& \textbf{ 0.41}& \textbf{ 0.39}& \textbf{ 0.39}& \textbf{ 0.39}& \textbf{ 0.39}& \textbf{ 0.39}& \textbf{ 0.39}& \textbf{ 0.41} \\ 
\textbf{AR(1)} \\ 
1 M & \textbf{ 0.49}&  0.49 &  0.47 &  0.47 &  0.46 & \textbf{ 0.46}& \textbf{ 0.46}& \textbf{ 0.44}& \textbf{ 0.44}& \textbf{ 0.39}& \textbf{ 0.40}& \textbf{ 0.40}& \textbf{ 0.40}& \textbf{ 0.37}& \textbf{ 0.39}& \textbf{ 0.37}& \textbf{ 0.39}& \textbf{ 0.40}& \textbf{ 0.39}& \textbf{ 0.40}& \textbf{ 0.40}& \textbf{ 0.40}& \textbf{ 0.42}& \textbf{ 0.40} \\ 
3 M &  0.40 &  0.40 &  0.40 &  0.38 &  0.38 &  0.36 &  0.38 &  0.40 &  0.40 &  0.40 &  0.40 &  0.42 &  0.40 &  0.42 &  0.42 &  0.42 &  0.42 &  0.45 &  0.47 &  0.47 &  0.47 &  0.47 &  0.47 &  0.47 \\ 
6 M &  0.31 &  0.33 &  0.35 &  0.33 &  0.33 &  0.33 &  0.33 &  0.40 &  0.40 &  0.40 &  0.40 &  0.40 &  0.40 &  0.40 &  0.40 &  0.40 &  0.40 &  0.40 &  0.42 &  0.42 & \textbf{ 0.42}& \textbf{ 0.44}& \textbf{ 0.44}& \textbf{ 0.46} \\ 
12 M & \textbf{ 0.17}& \textbf{ 0.17}& \textbf{ 0.17}& \textbf{ 0.17}& \textbf{ 0.22}& \textbf{ 0.22}& \textbf{ 0.22}& \textbf{ 0.22}& \textbf{ 0.22}& \textbf{ 0.26}& \textbf{ 0.28}& \textbf{ 0.35}& \textbf{ 0.39}& \textbf{ 0.41}& \textbf{ 0.41}& \textbf{ 0.41}& \textbf{ 0.48}& \textbf{ 0.52}& \textbf{ 0.50}& \textbf{ 0.57}& \textbf{ 0.61}& \textbf{ 0.63}& \textbf{ 0.67}& \textbf{ 0.67} \\ 
\textbf{VAR(1)} \\ 
1 M &  0.47 &  0.47 &  0.46 & \textbf{ 0.46}& \textbf{ 0.46}& \textbf{ 0.47}& \textbf{ 0.46}& \textbf{ 0.46}& \textbf{ 0.49}& \textbf{ 0.53}& \textbf{ 0.53}& \textbf{ 0.53}& \textbf{ 0.53}& \textbf{ 0.53}& \textbf{ 0.53}& \textbf{ 0.54}& \textbf{ 0.54}& \textbf{ 0.54}& \textbf{ 0.56}& \textbf{ 0.56}& \textbf{ 0.56}& \textbf{ 0.56}& \textbf{ 0.56}& \textbf{ 0.56} \\ 
3 M &  0.38 &  0.35 &  0.35 &  0.40 &  0.42 &  0.42 &  0.42 &  0.42 &  0.44 &  0.45 &  0.45 &  0.47 &  0.51 &  0.51 &  0.53 &  0.55 &  0.55 &  0.56 &  0.56 &  0.58 &  0.58 &  0.62 &  0.64 &  0.62 \\ 
6 M &  0.25 &  0.25 &  0.25 &  0.25 &  0.27 &  0.27 &  0.31 &  0.33 &  0.33 &  0.33 &  0.37 &  0.40 &  0.40 &  0.44 &  0.48 &  0.46 &  0.44 &  0.46 &  0.48 &  0.50 & \textbf{ 0.52}& \textbf{ 0.54}& \textbf{ 0.54}& \textbf{ 0.58} \\ 
12 M &  0.22 &  0.22 &  0.24 &  0.24 &  0.26 &  0.28 &  0.28 &  0.35 &  0.39 &  0.41 &  0.43 &  0.50 &  0.50 &  0.52 &  0.54 &  0.54 &  0.57 &  0.57 &  0.57 &  0.59 &  0.61 &  0.65 &  0.65 &  0.65 \\ 
\textbf{RW} \\ 
1 M & \textbf{ 0.47}&  0.46 & \textbf{ 0.46}& \textbf{ 0.49}& \textbf{ 0.49}& \textbf{ 0.49}& \textbf{ 0.47}& \textbf{ 0.47}& \textbf{ 0.47}& \textbf{ 0.49}& \textbf{ 0.47}& \textbf{ 0.47}& \textbf{ 0.49}& \textbf{ 0.49}& \textbf{ 0.53}& \textbf{ 0.53}& \textbf{ 0.49}& \textbf{ 0.51}& \textbf{ 0.49}& \textbf{ 0.49}& \textbf{ 0.49}& \textbf{ 0.51}& \textbf{ 0.51}& \textbf{ 0.53} \\ 
3 M &  0.47 &  0.49 &  0.49 &  0.47 &  0.47 &  0.47 &  0.47 &  0.49 &  0.49 &  0.49 &  0.49 &  0.51 &  0.49 &  0.45 &  0.45 &  0.47 &  0.49 &  0.47 &  0.47 &  0.49 &  0.49 &  0.49 &  0.49 &  0.49 \\ 
6 M &  0.40 &  0.40 &  0.42 &  0.46 &  0.50 &  0.50 &  0.50 &  0.52 &  0.52 &  0.52 &  0.52 &  0.52 &  0.54 &  0.54 &  0.58 &  0.58 &  0.58 &  0.58 &  0.58 &  0.60 & \textbf{ 0.60}& \textbf{ 0.60}& \textbf{ 0.60}& \textbf{ 0.60} \\ 
12 M & \textbf{ 0.41}& \textbf{ 0.41}& \textbf{ 0.41}& \textbf{ 0.41}& \textbf{ 0.43}& \textbf{ 0.43}& \textbf{ 0.50}& \textbf{ 0.50}& \textbf{ 0.54}& \textbf{ 0.57}& \textbf{ 0.59}& \textbf{ 0.63}& \textbf{ 0.63}& \textbf{ 0.61}& \textbf{ 0.61}& \textbf{ 0.61}& \textbf{ 0.63}& \textbf{ 0.63}& \textbf{ 0.63}& \textbf{ 0.63}& \textbf{ 0.67}&  0.70 &  0.70 &  0.70 \\ 
\bottomrule
    \end{tabular}}
    \caption{The table reports average number of cases when the error from forecasts of four competing models FTDNN, AR(1), VAR(1), and RW are positive. The higher the number from 0.5, the more over-predictions model yields. We consider multiple-step-ahead forecasts of 1,3,6, and 12 months for individual maturities. The Model Confidence Set (MSC) is used to compare the MME(O) errors across the four competing models. The ratio is bold in case of the model belonging to the $\widehat{\mathcal{M}}^*_{10\%}$.}
  \label{tab:MMEO}
\end{table}
\end{landscape}

\end{document}